\documentclass[preprint,review,12pt,leqno]{elsarticle}

\usepackage{float}
\usepackage{graphicx}
\usepackage{multirow}
\usepackage{pdflscape}
\usepackage{booktabs}
\usepackage{color,amsmath,amssymb}
	\usepackage{dcolumn}
	\usepackage{geometry}
	\usepackage{pifont}
	\usepackage{fleqn}
	\usepackage{hyperref}
	\usepackage{url}
	\usepackage[T1]{fontenc}
	\usepackage{ae,aecompl}
	\usepackage[normalem]{ulem}
	\usepackage{array}
	\usepackage{epsf}
	\usepackage{epstopdf}
	\usepackage{amssymb}
	\usepackage{longtable}
	\usepackage{epsfig}
	\usepackage{gensymb}

	\newcommand{\hcch}{C$_2$H$_2$ }
	
	\newcommand{\Marvel}{{\sc Marvel}}
	\newcommand{\lfour}{$\ell_4$}
	\newcommand{\lfive}{$\ell_5$}

	\usepackage{units}
	
	\usepackage[varg]{txfonts}
	\usepackage{verbatim}

\begin{document}

\title{\Marvel\ analysis of the measured high-resolution rovibrational spectra of \hcch}
	
\author{Katy L. Chubb$^{a1}$}
\author{Megan Joseph$^b$, Jack Franklin$^b$, Naail Choudhury$^b$}
\author{Tibor Furtenbacher$^c$, Attila G. Cs\'asz\'ar$^c$}
\author{Glenda Gaspard$^b$, Patari Oguoko$^b$, Adam Kelly$^b$}
\author{Sergei N. Yurchenko,$^a$
Jonathan Tennyson,$^a$\footnote{To whom correspondence should be addressed; email: katy.chubb.14@ucl.ac.uk, j.tennyson@ucl.ac.uk}}
\author{Clara Sousa-Silva.$^{d,a,b}$}
\address{$^a$Department of Physics and Astronomy, University College
London, London, WC1E 6BT, UK}
\address{$^b$Highams Park School, Handsworth Avenue, Highams Park,
London, E4 9PJ, UK}
\address{$^c$Institute  of Chemistry,
E\"otv\"os Lor\'and University and MTA-ELTE Complex Chemical
Systems Research Group,
H-1518 Budapest 112, Hungary}
\address{$^d$Department of Earth, Atmospheric and Planetary Sciences, Massachusetts Institute of Technology, 77 Massachusetts Ave,
Cambridge, MA 02139, USA}

\date{\today}	

\begin{abstract}
  Rotation-vibration energy levels are determined for the electronic
  ground state of the acetylene molecule, $^{12}$C$_2$H$_2$, using
  the Measured Active Rotational-Vibrational Energy Levels (\Marvel)
  technique. 37,813 measured transitions from 61 publications are
  considered. The distinct components of the spectroscopic network linking ortho and para states are
  considered separately. The 20,717 ortho and 17,096 para transitions
  measured experimentally are used to determine 6013 ortho and 5200
  para energy levels.  The \Marvel\ results are compared with
  alternative compilations based on the use of effective Hamiltonians.

\end{abstract}

\maketitle
\newpage

\section{Introduction}\label{sec:intro}

Acetylene, HCCH, is a linear tetratomic unsaturated hydrocarbon whose
spectrum is important in a large range of environments. The temperatures of these environments range
from the hot, oxy-acetyene flames which are widely used for welding
and related activities \cite{12Gaxxxx.C2H2}, temperate, where
monitoring of acetylene in breath gives insights into the nature of
exhaled smoke \cite{10MeScSk.C2H2}, to the cold, where the role of
acetylene in the formation of carbon dust in the interstellar medium
is a subject of debate \cite{14DhRaxx.C2H2}. Furthermore, acetylene is
observed in star-forming regions \cite{76RiHaKl.C2H2} and thought to
be an important constituent of clouds in the upper atmospheres of
brown dwarfs and exoplanets \cite{13BiRiHe.dwarfs}.  Acetylene
provides a major source of opacity in the atmospheres of cool
carbon stars \cite{82RiBaRa.C2H2,04GaHoJo}. It is present in various
planetary and lunar atmospheres in the solar system, including Jupiter
and Titan \cite{08OrVoxx.C2H2} and has been detected on comets
\cite{96BrToWe.C2H2}. The first analysis of the atmosphere of a
super-Earth, exoplanet 55 Cancri e \cite{jt629}, speculate that
acetylene could be present in its atmosphere; however the spectral
data currently available does not allow for an accurate verification
of its presence in such a high temperature environment.

The spectroscopy of acetylene has long been studied in the laboratory, particularly
by the group of Herman in Brussels. A full analysis of these experimental studies is
given below. Herman and co-workers have presented a number of reviews of the rovibrational
behaviour of acetylene in X~$^1\Sigma^+_g$ ground electronic state \cite{07Hexxxx.C2H2,10DiHexx.C2H2,11Herman.C2H2}.
Besides summarizing the status rotation-vibration spectroscopy of the system,
these reviews also give insight into the internal dynamics of the system, a topic
not considered here. 

From a theoretical point of view a number of variational nuclear motion
calculations have been performed for the acetylene ground electronic state
\cite{93BrHaxx.method,96Scxxxx.linear,jt346,02XuLiXi.C2H2,03XuGuZo.C2H2,jt479}.
New theoretical
rovibrational calculations for this molecule are in progress as part of the
ExoMol project~\cite{jt528,jt631}, a database of theoretical
line lists for molecules of astrophysical importance, appropriate up to high
temperatures of around 300 -- 3000K, for use in characterising the atmospheres of
cool stars and exoplanets. High accuracy experimental energy levels provide
essential input for testing and improving theoretically calculated line positions.

In this work we present the largest compilation of published
experimental data on rovibrational transitions for the acetylene
molecule, which has been formatted and analysed using the \Marvel\
(Measured Active Rotational-Vibrational Energy Levels) spectroscopic
network software, the results of which are presented and discussed in
this paper. The next section gives the underlying theory used for the
study. Section 3 presents and discusses the experimental sources used.
Results are given in Section 4. Section 5 discusses these results;
this section presents comparisons with recent empirical databases due
to Amyay {\it et al.}  \cite{16AmFaHe.C2H2} (henceforth 16AmFaHe),
Lyulin and Campargue \cite{17LyCaxx.C2H2} (henceforth 17LyCa) and
Lyulin and Perevalov, \cite{17LyPe.C2H2} (henceforth 17LyPe), which
builds on their earlier work \cite{16LyPexx.C2H2}, all of which
only became available while the present study was being undertaken. Finally
section 6 gives our conclusions.

\section{Theory}

\subsection{MARVEL}
The \Marvel\ procedure \cite{jt412,12FuCsxx.marvel} is based on the
theory of spectroscopic networks
\cite{11CsFuxx.marvel,16ArPeFu.marvel} and is principally based on earlier work by Flaud \textit{et al.} \cite{76FlCaMa} and Watson \cite{04Watson,94Watson}. The \Marvel\ program can be
used to critically evaluate and validate experimentally-determined
transition wavenumbers and uncertainties collected from the
literature. It inverts the wavenumber information to obtain accurate
energy levels with an associated uncertainty. \Marvel\ has been
successfully used to evaluate the energy levels for molecules, most
recently TiO \cite{jt672} and others such as $^{14}$NH$_3$
\cite{jt608,jtNH3update}, water vapour
\cite{jt454,jt482,jt539,jt576,jtwaterupdate}, H$_2$D$^+$ and
D$_2$H$^+$ \cite{13FuSzFa.marvel}, H$_3^+$ \cite{13FuSzMa.marvel}, and
C$_2$ \cite{jt637}. To be useful for \Marvel, measured transitions
must have an associated uncertainty and each state must be uniquely labelled, typically by a set of quantum numbers. It should be noted that while \Marvel\ requires
uniqueness it does not require these quantum numbers to be strictly correct,
or indeed even meaningful, beyond obeying rigorous selection rules; 
these
assignments simply act as labels for each state. Nevertheless, it greatly aids
comparisons with other data if they contain physically sensible information. The quantum
numbers used in the present study are considered in the following
section.

\subsection{Quantum number labelling}

The 11 quantum numbers that were used for labelling the upper and lower states
are detailed in Table~\ref{t:QN}. This includes the quanta of each vibrational mode in
normal mode notation: $v_1, v_2, v_3, v_4, l_4, v_5, l_5, K=|\ell_4+\ell_5|$
and $J$, where $v_1,\ldots,v_5$ are the vibrational quantum numbers, \lfour\ and
\lfive\ are the vibrational angular momentum quantum numbers associated with
$v_4$ and $v_5$, respectively, with $|\ell|=v,v-2\ldots1$ for odd $v$, $|\ell|=v,v-2\ldots$0 for even $v$. $K=|k|$ is the rotational quantum number, with $k$ corresponding to the projection of the rotational angular momentum, ${\bf J}$, on the $z$ axis. $K$ is also equal to the total vibrational angular momentum quantum number, $|L| = |\ell_4+\ell_5|$, and therefore $K$ will be also referred to as the total vibrational angular momentum. $J$ is the quantum number associated with rotational angular momentum, ${\bf J}$.  We follow the phase convention of the Belgium group, \cite{11Herman.C2H2} for $K\equiv|k|=|\ell_4+\ell_5|$ with $\ell_4\geq0$ if $k=0$. We also use the $e$ or $f$ labelling, along with the nuclear spin state (ortho or para).

\begin{table}
\caption{Quantum numbers used to label the upper and lower energy states.}
\label{t:QN}
\footnotesize
\begin{tabular}{ll}
\hline\hline
Label & Description\\
\hline
$v_1$	& CH symmetric stretch ($\sigma_g^{+}$) \\
$v_2$	& CC symmetric stretch ($\sigma_g^{+}$) \\
$v_3$	& CH antisymmetric stretch ($\sigma_u^{+}$) \\
$v_4$	& Symmetric (trans) bend ($\pi_g$) \\
$\ell_4$	& Vibrational angular momentum associated with $v_4$ \\
$v_5$	& Antisymmetric (cis) bend ($\pi_u$) \\
$\ell_5$	& Vibrational angular momentum associated with $v_5$ \\
$K$	& Total vibrational angular momentum, $|\ell_4+\ell_5|$ and Rotational quantum number\\
$J$	& Rotational angular momentum \\
$e/f$ & Symmetry relative to the Wang transformation (see text) \\
ortho/para	& Nuclear spin state (see text) \\
\hline\hline
\end{tabular}
\end{table}

The quantum number assignments for this work were taken from the original sources where possible, with any exceptions noted in section 3.1 and 3.2: particular reference should be made to the general comments (1a) and (1b) in 3.2. While \Marvel\ requires a unique set of quantum numbers for each state, it merely treats these as labels and whether they are strictly correct or not does not effect the validity of results. Nevertheless, the attempt to label them with sensible assignments aids comparisons with other datasets.

Levels with parity $+(-1)^J$ are called $e$ levels and those with parity $-(-1)^J$ are called $f$ levels. In other words, $e$ and $f$ levels transform in the same way as the rotational levels of $^1\Sigma^+$ and $^1\Sigma^-$  states, respectively \cite{75BrHoHu.linear}. Table \ref{t:parity} gives the combinations of $e/f$ and $J$ with corresponding parity. 
States of a linear molecular are often also classified based on inversion, with states which are left unchanged called `gerade' and labelled with a subscript $g$, and those whose phase changes to opposite are called `ungerade' and labelled $u$. The ortho and para labels  are defined based on the the permutation, $P$, of the identical hydrogen atoms. For the para states  the corresponding rovibrational wavefunctions, $\Psi_{\rm r-v}$, are symmetric, i.e. $P\,\Psi_{\rm r-v}=(+1)\Psi_{\rm r-v}$, while for the ortho states they are antisymmetric, $P\,\Psi_{\rm r-v}=(-1)\Psi_{\rm r-v}$.  The allowed combinations of these labels are shown in Table~\ref{t:labels} and explained in more detail below.

\begin{table}[H]
	\centering
	\caption{Parity of states in $^{12}$C$_2$H$_2$ based on the symmetry labels used in this work.}
	\label{t:parity}
	\begin{tabular}{ccc}
		\hline
		\hline
		$e/f$ & $J$ & Parity \\
		\hline
		$e $& Odd &$ - $  \\
		$e $& Even &$ + $ \\
		$f $& Odd &$ + $\\
		$f $& Even &$ - $ \\
		\hline
	\end{tabular}
\end{table}

\begin{table}[H]
	\centering
	\caption{Allowed combinations of symmetry labels for rovibrational states (including spin) of $^{12}$C$_2$H$_2$, where $s = $ symmetric, $a=$ antisymmetric, `Total' is how the rovibronic wavefunction, including the nuclear spin, acts under permutation symmetry.}
	\label{t:labels}
	\begin{tabular}{ccccc}
		\hline
		\hline
		$u/g$ & $+/-$ & Ro-vib. & Nuclear spin & Total  \\
		\hline
		$u $&$ + $&$a$ & Ortho & $a$ \\
		$u $&$ - $&$s$& Para  & $a$  \\
		$g $&$ + $&$s$& Para & $a$ \\
		$g $&$ - $&$a$& Ortho & $a$ \\
		\hline
	\end{tabular}
\end{table}

The $e/f$ labelling which has been adopted in this work was originally introduced by Brown {\it et al.} \cite{75BrHoHu.linear} in order to eliminate issues relating to Pl\'iva's  $c/d$ labelling \cite{72Plxxxxa.C2H2} and the $s/a$ labelling of Winnewisser and Winnewisser \cite{72WiWi.HCNO}. For more detailed information on the $e/f$ rotational splitting, see the section
titled `$e/f$ levels', page 173 of Herman {\it et al.} \cite{99HeLiVa.linear}. In summary, an
interaction known as $\ell$-doubling  occurs in linear molecules, which
splits the rotational, $J$, levels in certain vibrational states. The symmetry describing these states is based on the total vibrational angular
momentum quantum number, $K$. There are, for example, two distinct states in the 2$\nu_4$ band; one
with $K=0$ ($\Sigma_{g}^{+}$, $(0002^{0}0^{0})^0$) and the other $K=2$
($\Delta_g$, $(0002^{2}0^{0})^2$). In this case, the interaction with the rotation leads to a splitting  of the rovibrational levels in the $K=2$ ($\Delta_g$) sublevel ($\ell$-doubling). 
The $\Delta_e$ (corresponding to one
of the two bending modes) and $\Sigma_e$ (corresponding to one of the three
stretching modes) states repel each other, pushing $\Delta_e$ to a lower energy while $\Delta_f$ is unaffected. For this reason the $e$ state typically lies below the $f$ state, as bending occurs at a lower frequency than stretching \cite{99HeLiVa.linear}.
This effect is $J(J+1)$ dependent and so becomes increasingly important at higher
rotational excitations.  If a rovibrational state has no
rotational splitting (as is the case if both $\ell_4$=0 and $\ell_5=0$, but not if $\ell_4=1$ and $\ell_5=-1$), the state is always labelled $e$ and there is no
corresponding $f$ state.

Herman and Lievin \cite{82HeLixx.C2H2} give an excellent description of the ortho and para states
of acetylene which is summarised here. The hydrogen atoms in the main
isotopologue of acetylene are spin-$\frac{1}{2}$ particles and therefore, as
Fermions, obey Fermi-Dirac rules. The $^{12}$C carbon atoms, the only isotopologue considered in this work, are spin-0 and so do not need
to be considered here. The symmetry operation $P$ describes a permutation of
identical particles; when applied to the molecule it implies permutation of the
two hydrogen atoms. The total wavefunction must either be symmetric or
antisymmetric upon such a transformation. In the case of fermions it must be
antisymmetric.
The permutation symmetry of the ground electronic state is totally symmetric upon interchange of identical atoms and so the electronic part of the wavefunction can be ignored for this situation. The symmetry of the nuclear spin part of the
wavefunction is not usually specified, but can easily be deduced from the remaining symmetry.
If the rovibrational part of the wavefunction is antisymmetric under permutation symmetry (resulting from a combination of $g$ and $-$ or $u$ and $+$), then the nuclear spin state must be ortho and if the rovibrational part of the wavefunction is symmetric ($g,+$ or $u,-$), then the nuclear spin state must be para (see Table \ref{t:labels}).

It is important to distinguish the vibrational and rotational symmetries  from the symmetry of the rovibrational states of $\Psi_{\rm r-v}$. For  a  linear molecule such as $^{12}$C$_2$H$_2$ both the rotational $\Psi_{\rm r}$ and vibrational $\Psi_{\rm v}$ contributions to $\Psi_{\rm r-v}$  should transform according with the point group $D_{\infty h}(M)$, spanning an infinite number of irreducible representations such as $\Sigma_{g/u}^{+/-}$ ($K=0$), $\Pi_{g/u}^{+/-}$ ($K=1$), $\Delta_{g/u}^{+/-}$ ($K=2$) etc.  However, after combining the rotational and vibrational parts into the rovibrational state $\Psi_{\rm r-v}$, only the $K=0$ states (i.e. $\Sigma_{g}^{+}$, $\Sigma_{g}^{-}$, $\Sigma_{u}^{+}$, $\Sigma_{u}^{-}$) can lead to the total nuclear-rotation-vibrational state obeying the proper nuclear statistics, as described above. These are the irreducible elements of the $D_{2h}$(M) group  \citep{98BuJexx.method}, which  according to our labeling scheme correspond to the four pairs: $e$~ortho, $e$~para, $f$~ortho and $f$~para.
For example  the vibrational state $\nu_5$ ($\Pi_{u}$)  can be combined with the $J=1,K=1$ ($\Pi_g$) rotational state to produce three rovibrational combinations of $\Sigma_{u}^{+}$, $\Sigma_{u}^{-}$ and $\Pi_{u}$ ($D_{\infty h}$ point group). However only the $\Sigma_{u}^{-}$, $\Sigma_{u}^{+}$ states are allowed by the nuclear statistics. Here $\nu_5$, $\Pi_{u}$, $K$, $\Pi_{g}$ are not rigourous quantum numbers/labels, while $J=1$, $e/f$ and ortho/para are. Thus these two rovibrational states are assigned $(0000^01^1)^1, $J$=1, e$, para and $(0000^01^1)^1, $J=$1, f$, ortho, respectively.
It should be also noted that generally neither $K$ nor $v_1,\ldots,v_5$ are good quantum numbers. However the quantity $(-1)^{v_3+v_5}$ is s good quantum number as it defines the conserved $u/g$  symmetry as follows: a state is ungerade if $(-1)^{v_3+v_5}=-1$ and gerade if $(-1)^{v_3+v_5} =1$. The $+/-$ labelling is derived from $e/f$ and $J$, as given in Table \ref{t:parity}.

Throughout this paper we shall use the notations $(v_1 v_2 v_3 v_4^{\ell_4} v_5^{\ell_5})^K$ to describe vibrational states and $(v_1 v_2 v_3 v_4^{\ell_4} v_5^{\ell_5})^K$, $J$, $e/f$, ortho/para to describe rovibrational states. The $e$ and $f$ labelling combined with $J$ and nuclear spin state (ortho or para) gives the rigorous designation of each state. Other quantum number labels
are approximate but, besides representing the underlying physics, are
necessary to uniquely distinguish each state. The symmetry labels of the vibrational states ($\Sigma_{u/g}^{+/-}$, $\Pi_{u/g}$, $\Delta_{u/g}$, \ldots) have been added to the end of the output energy files (see Table~\ref{t:MARVEL} and supplementary material).

\subsection{Selection rules}

The rigourous selection rules governing rotation-vibration transitions for a symmetric linear molecule (molecular group D$_{{\infty}h}$(M)) are given by
\begin{align}
\Delta J =\pm 1 & \quad  \text{ with } \quad e{\leftrightarrow}e \quad  \text{ or } \quad f{\leftrightarrow}f,\\
\Delta J =\pm 0 & \quad   \text{ with } \quad  e{\leftrightarrow}f \\
J'+J'' \ne  0 \\
u{\leftrightarrow}g
\end{align}
The first two equations here correspond to the standard selection rule $+ \leftrightarrow -$ for the dipole transitions in terms of the parities. The ortho states of $^{12}$C$_2$H$_2$ have the statistical weight $g_{\rm ns}=3$, while for the para states $g_{\rm ns}=1$.

\section{Experimental sources}

A large number of experimentally determined transition frequencies can
be found in the literature for the main isotopologue of acetylene,
$^{12}$C$_2$H$_2$.  As part of this study we attempted to conduct
a rigorous and comprehensive search for all useable spectroscopic
data. This includes the transition frequency (in cm$^{-1}$) and
associated uncertainty, along with quantum number assignments for both
the upper and lower energy states. A unique reference label is
assigned to each transition, which is required for \Marvel\ input.
This reference indicates the data source, table (or page) and line
number that the transition originated from. The data source tag is
based on the notation employed by the IUPAC Task Group on water
\cite{jt482,jt562} with an adjustment discussed below. The associated
uncertainities were taken from the experimental data sources where
possible, but it was necessary to increase many of these in order to
achieve consistency with the same transition in alternative data
sources. As noted by Lyulin and Perevalov \cite{16LyPexx.C2H2}, these
sources often provide overall uncertainties for the strongest lines in
a vibrational band which may underestimate the uncertainty associated
with some or all of
the weaker, and especially of blended, lines.

61 sources of experimental data were considered. Two of the data
compilations mentioned in the introduction
\cite{16AmFaHe.C2H2,17LyCaxx.C2H2} contain data from multiple other
sources, some of which was not directly available to us.  Data taken
from these compilations is given a tag based on that used in the
compilation with the original reference given in Table~\ref{t:source}.  After
processing, 60 sources were used in the final data set. The data from
more recent papers is generally provided in digital format, but some
of the older papers had to be processed through digitalisation
software, or even manually entered in the most extreme cases. After
digitalisation the data was converted to \Marvel\ format; an example of
the input file in this format is given in Table \ref{t:ortho}; the full file can
be found in the supplementary data of this paper.

\begin{table}
	\caption{Extract from the \Marvel\ input file for the ortho transitions. The full file is supplied as part of the supplementary information to this paper. All energy term values
and uncertainties are in units of cm$^{-1}$. The assignments are detailed in Table~\ref{t:QN}.
\label{t:ortho}
}\tt\footnotesize
\begin{tabular}{rccccr}
	\hline\hline
Energy & Unc & Upper assignment & Lower assignment & Ref  \\
	\hline
1248.2620 & 0.0005 & 0 0 0 1 1 1 -1 0 34 e ortho & 0 0 0 0 0 0 0 0 35 e ortho & 00Vander\_table2\_l1 \\
1252.8546 & 0.0005 & 0 0 0 1 1 1 -1 0 32 e ortho & 0 0 0 0 0 0 0 0 33 e ortho & 00Vander\_table2\_l2 \\
1257.4230 & 0.0005 & 0 0 0 1 1 1 -1 0 30 e ortho & 0 0 0 0 0 0 0 0 31 e ortho & 00Vander\_table2\_l4 \\
1261.9694 & 0.0005 & 0 0 0 1 1 1 -1 0 28 e ortho & 0 0 0 0 0 0 0 0 29 e ortho & 00Vander\_table2\_l6 \\
1266.4970 & 0.0005 & 0 0 0 1 1 1 -1 0 26 e ortho & 0 0 0 0 0 0 0 0 27 e ortho & 00Vander\_table2\_l8 \\
1271.0098 & 0.0005 & 0 0 0 1 1 1 -1 0 24 e ortho & 0 0 0 0 0 0 0 0 25 e ortho & 00Vander\_table2\_l10 \\
1275.5122 & 0.0005 & 0 0 0 1 1 1 -1 0 22 e ortho & 0 0 0 0 0 0 0 0 23 e ortho & 00Vander\_table2\_l11 \\
	\hline
\end{tabular}
\end{table}

Table~\ref{t:source} gives a summary of all the data sources used in this work,
along with the wavelength range, number of transitions, number of vibrational bands, the
approximate temperature of the experiment and comments, which can be
found in section 3.1.  Table~\ref{t:consid} gives those data sources which were considered
but not used, with comments on the reasons. The reference label given
in these tables corresponds to the unique labels in the \Marvel\ input
files, given in the supplementary data and illustrated in the last
column of Table \ref{t:ortho}. As transitions do not occur between ortho and para
states, they form two completely separate components of the experimental spectroscopic network, with
no links between them. All input and output files supplied in the
supplementary data to this work are split into either ortho or para.

{\footnotesize
\begin{longtable}{llcccrl}
\caption{Data sources used in this study with wavelength range, numbers of transitions and approximate temperature of the experiment. A/V stands for the number of transitions analysed/verified. 'RT' stands for room temperature. See section 3.1 for the notes.  
\label{t:source}
} \\
\hline
	\cr Tag &	Ref.	&	Range (cm$^{-1}$)	&	A/V & Bands & Temp & Note \\
	\hline\hline
	\endhead
	09YuDrPe	&	\cite{09YuDrPe.C2H2}	&	29-55	&	20/20	&	5	&	RT	&	\\	
	16AmFaHe\_kab91	&	\cite{91KaHeDi.C2H2}	&	61-1440	&	3233/3233	&	47	&	RT	&	\\	
	16AmFaHe\_amy10	&	\cite{10AmHeFa.C2H2}	&	63-7006	&	1232/1232	&	36	&	RT	&	\\	
	11DrYu	&	\cite{11DrYuxx.C2H2}	&	85-92	&	20/20	&	7	&	RT	&	\\	
	17JaLyPe	&	\cite{17JaLyPe.C2H2}	&	429-592	&	627/627	& 9	&	RT	&	\\	
	81HiKa	&	\cite{81HiKaxx.C2H2}	&	628-832	&	684/684	&	5	&	RT	&	(3a)	\\
	93WeBlNa	&	\cite{93WeBlNa.C2H2}	&	632-819	&	1610/1609	&	13	&	RT	&	(3b)	\\
	00MaDaCl	&	\cite{00MaDaCl.C2H2}	&	644-820	&	77/77	&	1	&	RT	&	\\	
	01JaClMa	&	\cite{01JaClMa.C2H2}	&	656-800	&	355/355	&	4	&	RT	&	\\	
	50BeNi	&	\cite{50BeNixx.C2H2}	&	671-4160	&	500/0	&	13	&	RT	&	(3c)	\\
	16AmFaHe\_gom10	&	\cite{10GoJaLa.C2H2}	&	1153-1420	&	27/27	&	3	&	RT	&	\\	
	16AmFaHe\_gom09	&	\cite{09GoJaLa.C2H2}	&	1247-1451	&	66/66	&	8	&	RT	&	\\	
	00Vander	&	\cite{00Vander.C2H2}	&	1248-1415	&	64/64	&	2	&	RT	&	\\	
	16AmFaHe\_amy09	&	\cite{09AmRoHe.C2H2}	&	1253-3422	&	3791/3777	&	57	&	Up to 1455K 	&	(3d)	\\
	03JaMaDa	&	\cite{03JaMaDab.C2H2}	&	1810-2235	&	486/486	&	14	&	RT	&	\\	
	03JaMaDab	&	\cite{03JaMaDa.C2H2}	&	3207-3358	&	109/109	&	2	&	RT	&	\\	
	16AmFaHe\_jac02	&	\cite{02JaMaDa.C2H2}	&	1860-2255	&	150/150	&	3	&	RT	&	\\	
	72Pliva	&	\cite{72Plxxxxa.C2H2}	&	1865-2598	&	1016/1015	&	15	&	RT	&	\\	
	16AmFaHe\_ber98	&	\cite{98BeCaLo.C2H2}	&	1957-1960	&	19/19	&	1	&	RT	&	(3e)	\\
	16AmFaHe\_jac07	&	\cite{07JaLaMa.C2H2}	&	2515-2752	&	148/148	&	3	&	RT	&	\\	
	16AmFaHe\_pal72	&	\cite{72PaMiNa.C2H2}	&	2557-5313	&	42/42	&	3	&	RT	&	\\	
	16AmFaHe\_vda93	&	\cite{93VaHuCa.C2H2}	&	2584-3364	&	499/499	&	5	&	RT	&	\\	
	93DcSaJo	&	\cite{93DcSaJo.C2H2}	&	2589-2760	&	372/372	&	3	&	RT	&	\\	
	82RiBaRa	&	\cite{82RiBaRa.C2H2}	&	3140-3399	&	1789/1788	&	21	&	RT and 433K 	&	\\	
	16AmFaHe\_sarb95	&	\cite{95SaDcGub.C2H2}	&	3171-3541	&	401/401	&	8	&	RT	&	\\	
	06LyPeMa	&	\cite{06LyPeMa.C2H2}	&	3182-3327	&	167/167	&	13	&	RT	&	\\	
	16AmFaHe\_man05	&	\cite{05MaJaDa.C2H2}	&	3185-3355	&	288/288	&	5	&	RT	&	\\	
	16AmFaHe\_sara95	&	\cite{95SaDcGu.C2H2}	&	3230-3952	&	424/424	&	5	&	RT	&	\\	
	16AmFaHe\_ber99	&	\cite{99BeMaLo.C2H2}	&	3358-3361	&	21/21	&	1	&	RT	&	(3e)	\\
	16AmFaHe\_lyub07	&	\cite{07LyPeMa.C2H2}	&	3768-4208	&	668/668	&	8	&	RT	&	\\	
	16AmFaHe\_gir06	&	\cite{06GiFaSo.C2H2}	&	3931-4009	&	91/91	&	10	&	RT	&	\\	
	16AmFaHe\_dcu91	&	\cite{91DcSaGu.C2H2}	&	3999-4143	&	251/251	&	6	&	RT	&	\\	
	72BaGhNa	&	\cite{72BaGhNa.C2H2}	&	4423-4791	&	472/408	&	8	&	RT	&	(3f)	\\
	16AmFaHe\_lyua07	&	\cite{07LyPeGu.C2H2}	&	4423-4786	&	440/440	&	8	&	RT	&	\\	
	16AmFaHe\_lyu08	&	\cite{08LyJaLa.C2H2}	&	5051-5562	&	320/320	&	7	&	RT	&	\\	
	16AmFaHe\_kep96	&	\cite{96KeMeKl.C2H2}	&	5705-6862	&	1957/1957	&	30	&	RT	&	\\	
	17LyCa	&	\cite{17LyCaxx.C2H2}	&	5852-8563	&	4941/4941	&	108	&	RT	&	(3g)	\\
	16AmFaHe\_rob08	&	\cite{08RoHeFa.C2H2}	&	5885-6992	&	568/568	&	20	&	RT	&	\\	
	07TrMaDa	&	\cite{07TrMaDa.C2H2}	&	6299-6854	&	546/546	&	13	&	RT	&	(3h)	\\
	16AmFaHe\_lyu09	&	\cite{09LyPeTr.C2H2}	&	6300-6666	&	89/89	&	5	&	RT	&	\\	
	16KaNaVa	&	\cite{16KaNaVa.C2H2}	&	6386-6541	&	19/19	&	2	&	RT	&	(3i)	\\
	16AmFaHe\_kou94	&	\cite{94KoGuTe.C2H2}	&	6439-6629	&	73/73	&	1	&	RT	&	\\	
	15TwCiSe	&	\cite{15TwCiSe.C2H2}	&	6448-6564	&	135/135	&	2	&	RT	&	\\	
	02HaVa	&	\cite{02HaAuxx.C2H2}	&	6448-6685	&	271/271	&	4	&	RT	&	\\	
	77BaGhNa	&	\cite{77BaGhNa.C2H2}	&	6460-6680	&	860/859	&	15	&	RT	&	(3j)	\\
	05EdBaMa	&	\cite{05EdBaMa.C2H2}	&	6472-6579	&	41/41	&	1	&	RT	&	\\	
	13ZoGiBa	&	\cite{13ZoGiBa.C2H2}	&	6490-6609	&	37/37	&	1	&	RT	&	\\	
	00MoDuJa	&	\cite{00MoDuJa.C2H2}	&	6502-6596	&	36/36	&	1	&	RT	&	\\	
	96NaLaAw	&	\cite{96NaLaAw.C2H2}	&	6502-6596	&	36/36	&	1	&	RT	&	\\	
	16AmFaHe\_amy11	&	\cite{11AmHeFa.C2H2}	&	6667-7868	&	2259/2256	&	79	&	RT	&	(3k)	\\
	15LyVaCa	&	\cite{15LyVaCa.C2H2}	&	7001-7499	&	2471/2471	&	29	&	RT	&	(3l)\\	
	09JaLaMa	&	\cite{09JaLaMa.C2H2}	&	7043-7471	&	233/233	&	4	&	RT	&	\\	
	02VaElBr	&	\cite{02VaElBr.C2H2}	&	7062-9877	&	626/626	&	11	&	RT	&	(3m)	\\
	16LyVaCa	&	\cite{16LyVaCa.C2H2}	&	8283-8684	&	627/627	&	14	&	RT	&	(3n)	\\
	17BeLyHu	&	\cite{17BeLyHu.C2H2}	&	8994-9414	&	432/432	&	11	&	RT	&	\\	
	89HeHuVe	&	\cite{89HeHuVe.C2H2}	&	9362-10413	&	657/657	&	14	&	RT	&	(3o)	\\
	93SaKa	&	\cite{93SaKaxx.C2H2}	&	12428-12538	&	91/73	&	1	&	RT	&	(3p)\\	
	03HeKeHu	&	\cite{03HeKeHu.C2H2}	&	12582-12722	&	60/60	&	1	&	RT	&	\\	
	92SaKa	&	\cite{92SaKaxx.C2H2}	&	12904-13082	&	216/212	&	3	&	RT	&	(3q)	\\
	94SaSeKa	&	\cite{94SaSeKa.C2H2}	&	13629-13755	&	53/53	&	1	&	<RT (223K)	&	(3r)	\\
	\hline
Total&		&	29-13755	& 37813/37206 & & & \\
	\hline\hline
\end{longtable}
}

{
\footnotesize
\begin{longtable}{lll}
	\caption{Data sources considered but not used in this work. \label{t:consid}}\\
	\hline\hline
\cr Tag	&	Ref.	&	Comments	\\
\hline
	16AmFaHe\_abb96	&	\cite{96TeHeSo.C2H2}	&	0 transitions in 16AmFaHe; data not available in original paper.\\
	16AmFaHe\_eli98	&	\cite{99ElLiCa.C2H2}	&	0 transitions in 16AmFaHe; data not available in original paper.\\
	72Plivaa	&	\cite{72Plxxxxb.C2H2}:	&	Energy levels only	\\
	02MeYaVa	&	\cite{02MeYaVa.C2H2}	&	No suitable data	\\
	01MeYaVa	&   \cite{01MeYaVa.C2H2}	&	No suitable data	\\
	99SaPeHa	&	\cite{99SaPeHa.C2H2}	&	No suitable data	\\
	97JuHa	&	\cite{97JuHaxx.C2H2}	&	No suitable data	\\
	93ZhHa	&	\cite{93ZhHaxx.C2H2}	&	No suitable data	\\
	93ZhVaHa	&	\cite{93ZhVaHa.C2H2}	&	No suitable data	\\
	91ZhVaKa	&	\cite{91ZhVaKa.C2H2}	&	No suitable data    \\
	13SiMeVa	&	\cite{13SiMeVa.C2H2}	&	No suitable data	\\
	83ScLeKl	&	\cite{83ScLeKl.C2H2}	&	No assignments given	\\
	\hline
\end{longtable}
}

\subsection{Comments on the experimental sources in Table~\protect\ref{t:source}}

\noindent
	\textbf{(3a)}  81HiKa \cite{81HiKaxx.C2H2} has an apparent misprint in column 2
of their Table 6: the R(19) line should be 780.2601 cm$^{-1}$ not 790.2601 cm$^{-1}$, as confirmed by 01JaClMa
\cite{01JaClMa.C2H2}, and in column 5 of their Table 4: the Q(3) line should be 728.9148 cm$^{-1}$ not
729.9148 cm$^{-1}$, also confirmed by 01JaClMa \cite{01JaClMa.C2H2}.\\
	\textbf{(3b)} 93WeBlNa\_page14\_l38 from 93WeBlNa \cite{93WeBlNa.C2H2} is not consistent with other data sources. It was marked in the original dataset as a transition that the authors did not include in their analysis and so has been removed from our dataset.\\
	\textbf{(3c)} 50BeNi \cite{50BeNixx.C2H2} was deemed too unreliable to use in
the final dataset: data are directly contradicted by other sources. \\
\textbf{(3d)} Many of the transitions included from 16AmFaHe\_amy09 \cite{09AmRoHe.C2H2} are not duplicated in any other source. While this means they represent a valuable source of data, and have thus been kept in the \Marvel\ dataset, the fact that there is no other experimental data to back them up means they should be treated with some degree of caution. As stated in the original paper, modelling such a high temperature region is a challenge. There are a small number of transitions - 14 out of 3791 - that do not match those from other data sources and have been removed from our dataset.\\
\textbf{(3e)} 16AmFaHe\_ber98 \cite{98BeCaLo.C2H2} and 16AmFaHe\_ber99 \cite{99BeMaLo.C2H2} are Raman spectra and so the transitions do not follow the selection rules detailed in section 2.3 of this paper. \\
\textbf{(3f)}  72BaGhNa \cite{72BaGhNa.C2H2} has a band labelled $(0013^10^0)^1$ - $(0001^10^0)^1$ which is not consistent with other data sources. It was found that the band labelled $(0104^01^1)^1$ - $(0001^10^0)^1$ gave energies consistent with those labelled  $(0013^10^0)^1$ - $(0001^10^0)^1$ in other data sources (16AmFaHe\_lyua07, 16AmFaHe\_lyu08). Bands including $(0104^01^1)^1$ are not present in other data sources. We have swapped the labelling of these bands accordingly. All other bands from this dataset were included, with the exception of the single transition labelled 72BaGhNa\_table2\_c2\_l32, which was not consistent with other datasets. \\
\textbf{(3g)}  17LyCa \cite{17LyCaxx.C2H2} provides a collection of data recorded in Grenoble
using cavity ring down spectroscopy (CRDS) from several
papers. 15LyVaCa (FTS15 in the notation of 17LyCa) \cite{15LyVaCa.C2H2}, 16LyVaCa (FTS16) \cite{16LyVaCa.C2H2} and 17BeLyHu (FTS17) \cite{17BeLyHu.C2H2} were all already
included as separate files in our dataset and so were removed from the 17LyCa
\cite{17LyCaxx.C2H2} dataset. The remaining data, CRDS13 \cite{13LyCaMo.C2H2},
CRDS14 \cite{14LyMoBe.C2H2} and CRDS16 \cite{16KaLyBe.C2H2} are all included
in the final dataset with the tag '17LyCa'. See also comment (3l). \\
\textbf{(3h)} 07TrMaDa \cite{07TrMaDa.C2H2} contains a band labelled $2\nu_2+(\nu_4+3\nu_5)^0_+$. $\ell_4$ and $\ell_5$ were assigned as 1 and -1 respectively, to be consistent with the labelling of 16AmFaHe\_kep96.\\
\textbf{(3i)} Full data for 16KaNaVa \cite{16KaNaVa.C2H2} was provided in
digital format from the corresponding author (private communication, Juho Karhu). \\
\textbf{(3j)} 77BaGhNa\_table3\_l205 of 77BaGhNa \cite{77BaGhNa.C2H2} does not fit with the same transition in two other sources. \\
\textbf{(3k)} 16AmFaHe\_amy11 \cite{11AmHeFa.C2H2} includes a band ($(1000^06^6)^6$ - $(0000^00^0)^0$ ) which has transitions from $J=0$ to $J=10,11,12$. These are not physical and so have been removed from the dataset. There is one other transition which we have removed which we have found to be inconsistent with the other datasets. \\
\textbf{(3l)} There has been some changes in the authors approach to labelling levels between 15LyVaCa \cite{15LyVaCa.C2H2} and 17LyCa \cite{17LyCaxx.C2H2}, see comment (3g) (Alain Campargue, private communication). This was partly to allow all bands to have unique labelling, as duplicate labels were provided in 15LyVaCa as indicated by $^{**}$ or $^{*}$ superscripts. We have relabelled these bands to fit with other data sources, for example 16AmFaHe\_amy11 \cite{11AmHeFa.C2H2}. We have been informed by the authors of 17LyCa that they are currently making amendments to their published dataset (Alain Campargue, private communication). Table~\ref{t:label} summarises the changes in labelling between 15LyVaCa, the current version of 17LyCa\_FTS15 (see supplementary data of \cite{17LyCaxx.C2H2}) and this work. \\
\textbf{(3m)} 02VaElBr \cite{02VaElBr.C2H2} is missing one band labelling in the footnote to
their Table 3. The missing label for the penultimate level is I = ($v_1$$v_2$$v_3$$v_4$$^{l_4}$$v_5$$^{l_5}$)$^K$ = (0020$^0$1$^1$)$^1$. Full data for this source was provided in digital format by the corresponding author (Jean Vander Auwera, private communication).\\
\textbf{(3n)}  16LyVaCa \cite{16LyVaCa.C2H2} has duplicate lines in the
(1110$^0$0$^0$)$^0$ band. Those which are inconsistent with other
sources were removed and thus not included in the final data set for the \Marvel\ analysis. It is possible that
they should be re-assigned. \\
\textbf{(3o)} The assignments given for the band labelled $(0122^02^0)^0$ - $(0000^00^0)^0$ in 89HeHuVe \cite{89HeHuVe.C2H2} require the upper state to have the parity of an f-level, which is unphysical if both $\ell_4$=0 and $\ell_5$=0. There can be no $e/f$ splitting in this case. We assumed this upper state should be labelled $(0122^22^{-2})^0$. We have amended and included these reassigned transitions in our dataset.\\
\textbf{(3p)} Table 1 of 93SaKa \cite{93SaKaxx.C2H2} has duplicates for the $e{\leftrightarrow}e$ transitions in the $(2021^10^0)^1$ - $(0000^01^1)^1$  vibrational band. Those which are inconsistent with other
sources were removed and thus not included in the final data set.\\
\textbf{(3q)} 92SaKa \cite{92SaKaxx.C2H2} contains some duplicate lines which have been assigned identical quantum numbers. Those which are inconsistent with other
sources were removed and thus not included in the final data set.\\
\textbf{(3r)}  94SaSeKa \cite{94SaSeKa.C2H2} gives two tables of data but only one
is assigned with vibrational quantum numbers, so data from the other table were not considered in this study. \\

{\footnotesize
	\begin{longtable}{lll}
		\caption{Changes in labelling between 15LyVaCa \cite{15LyVaCa.C2H2}, 17LyCa\_FTS15 \cite{17LyCaxx.C2H2} and this work, in the form ($v_1$$v_2$$v_3$$v_4$$^{\ell_4}$$v_5$$^{\ell_5}$)$^K$. See comment (3l) in the text.\label{t:label}} \\
		\hline\hline
		\cr 15LyVaCa & 17LyVa\_FTS15	&	This work	\\
		\hline
		$(0204^21^{-1})^{1**}$	&	$(0113^10^0)^1$		&	$(0204^11^0)^1$	\\
		$(0113^10^0)^1$		&	$(0204^01^1)^1$		&	$(0113^10^0)^1$	\\
		$(1102^01^1)^1$		&   $(1102^01^1)^1$		&	$(1102^11^0)^1$		\\
		$(1102^21^{-1})^{1**}$	&	$(0202^23^{-1})^1$	&	$(1102^01^{1})^1$	\\
		$(1102^21^{-1})^{1*}$	&	$(1102^21^{-1})^{1}$	&	$(1102^21^{-1})^{1}$	\\
		\hline
	\end{longtable}
}

\subsection{General comments}
A number of general issues had to be dealt with before consistent networks could
be obtained.

\textbf{(1a)} 16AmFaHe \cite{16AmFaHe.C2H2} released a collation and
analysis of experimental data in the middle of our collation and
analysis stage. The entire database was formatted into \Marvel\ format
so it could subsequently be run through the software and combined with
the other experimental sources referenced in this paper. Some of the
experimental sources featured in the 16AmFaHe database paper had
already been collated and formatted to \Marvel\ format prior to its
publication. These are 03JaMaDa \cite{03JaMaDab.C2H2}, 91KaHeDi
\cite{91KaHeDi.C2H2}, 06LyPeMa \cite{06LyPeMa.C2H2}, 07LyPeGu
\cite{07LyPeGu.C2H2}, 82RiBaRa \cite{82RiBaRa.C2H2}, 02VaElBr
\cite{02VaElBr.C2H2} and 00MoDuJa \cite{00MoDuJa.C2H2}.  We used a
\Marvel\ format version of 16AmFaHe's compilation to compare to our
data, as a further check to validate data had been digitised and
formatted correctly; the versions included in the present study come
from the original datasets for these papers. A few of the sources that
were cited in 16AmFaHe
were not included in our final dataset. There were 0 transitions in 16AmFaHe from \cite{96TeHeSo.C2H2} (abb96), \cite{99ElLiCa.C2H2} (eli98) or \cite{11DrYuxx.C2H2} (drou11). The data for \cite{11DrYuxx.C2H2} was taken from the original paper (see 11DrYu in Table~\ref{t:source}), but there was no data obviously available in the original papers for the other two sources. We have tried to keep the quantum number labelling consistent with that of 16AmFaHe as much as possible (see comment (1b) for an exception). Some other sources were labelled in order to make them consistent, in particular those cases were $\ell_4$ and $\ell_5$ were not defined in the original source. \\
\textbf{(1b)} Many of the \lfour\ and \lfive\ assignments were
inconsistent between different sources, were not given in the original
data (often only the total $K=|\ell_4+\ell_5|$ is given) or were
inconsistent between data in the same dataset. Examples include the bands
with upper energies labelled
$(v_1v_2v_3v_4^{\ell_4}v_5^{\ell_5})^K$ =
$(0002^{*}1^{*})^1$, $(1102^{*}1^{*})^1$ or $(0102^{*}1^{*})^1$ in
16AmFaHe. Using simple combination differences, with the known lower
value and given transition wavenumber, there was found to be more than
one value for the upper energy. We assume this duplication of quantum
numbers for different states is down to the different method of
analysis used in 16AmFaHe, which does not require a completely unique
set.  For example, for the upper level $(1102^{2}1^{-1})^1$, $J$=2, $e$,
there are two transitions which give as upper energy level of 7212.93
cm$^{-1}$ (from 16AmFaHe\_kep96) and three that give 7235.29 cm$^{-1}$
(from 16AmFaHe\_vda02 and 16AmFaHe\_rob08).
These same two energies can be found in multiple other sources (07TrMaDa, 15LyVaCa, 77BaGhNa, 02VaElBr), but the $\ell_4$ and $\ell_5$ assignment was inconsistent for states of the same upper energy. The decision was made to batch them together and assign the first energy level (7212.94 cm$^{-1}$ in this example) as $(1102^{2}1^{-1})^1$ and the second (7235.29 cm$^{-1}$ in this example) as $(1102^{0}1^{1})^1$. The same logic was applied to other bands with $K=|\ell_4+\ell_5|$=1.\\
\textbf{(1c)} The $e/f$ notation (see section 2.2) was mostly
specified in experimental papers, but some required additional
investigation in order to assign them in such a way as to be
consistent with other papers. The c/d notation in
\cite{72Plxxxxa.C2H2}, for example, is analogous to the $e/f$
notation used in this work. \\
\textbf{(1d)} All transitions which were considered but not processed
in the final dataset are labelled with \_ct at the end of the
reference and have a minus sign in front of the transition frequency,
at the start of the file. \Marvel\
software ignores any line with a negative wavenumber.\\

\subsection{Other comments}
The following are sources of the acetylene data in the HITRAN database (\cite{03JaMaDa.C2H2,HITRAN2004,HITRAN2008,HITRAN2012}): 16AmFaHe\_gom09 \cite{09GoJaLa.C2H2}, 16AmFaHe\_gom10 \cite{10GoJaLa.C2H2}, 96NaLaAw  \cite{96NaLaAw.C2H2} , 05EdBaMa \cite{05EdBaMa.C2H2}, 16AmFaHe\_lyua07 \cite{07LyPeGu.C2H2}, 16AmFaHe\_jac07  \cite{07JaLaMa.C2H2}, 16AmFaHe\_jac09 \cite{09JaLaMa.C2H2}, 00Vander \cite{00Vander.C2H2}, 02HaVa \cite{02HaAuxx.C2H2}, 03JaMaDab \cite{03JaMaDa.C2H2}, 16AmFaHe\_kab91 \cite{91KaHeDi.C2H2}, 72Pliva \cite{72Plxxxxa.C2H2}, 03JaMaDa \cite{03JaMaDab.C2H2}, 82RiBaRa \cite{82RiBaRa.C2H2}, 16AmFaHe\_vda93
\cite{93VaHuCa.C2H2}.

\section{Results}

The MARVEL website
(\url{http://kkrk.chem.elte.hu/marvelonline/marvel\_full.php}) has a version
of \Marvel\ which can be run online. The variable NQN (number of quantum
numbers) is 11 in the case of acetylene, given in Table~\ref{t:QN}. These are
required for both the lower and upper levels, as illustrated in Table~\ref{t:ortho}.

All energies are measured from the zero point energy (ZPE). This is
the energy of the ground rovibrational state, which is given a
relative energy of 0 and is included in the para set of energy levels.
The ortho set of energies therefore needs a `magic number' to be added
to all the \Marvel\ ortho-symmetry energies. Here the magic number was
taken as the ground vibrational $(0000^00^0)^0$, $J=1$ state of
16AmFaHe \cite{16AmFaHe.C2H2} who determined the value 2.3532864 cm$^{-1}$,
see Table~\ref{t:compare} below.  The output for the ortho
energies in the supplementary data, and the extract of the output file
in Table~\ref{t:MARVEL}, all have this magic number added. The para component of the spectroscopic network does not require a magic number as it contains the ground
rovibrational level, $(0000^00^0)^0$, $J=0$. There are a small number
(284 for ortho and 119 for para) of energy levels which are not joined
to the two principal components (PCs) of the network. If more experimental transitions became available
in the future it would be possible to link these to the PCs.

\begin{table}
\caption{Extract from the \Marvel\ output file for the ortho transitions. The full file is supplied as part of the supplementary information to this paper. All energies and uncertainties are in units of cm$^{-1}$. The assignments are detailed in Table~\ref{t:QN}.}
\tt \footnotesize
\setlength\tabcolsep{0.9ex}
\label{t:MARVEL}
\begin{tabular}{rrrrrrrrrrrccccr}
	\hline\hline
	\multicolumn{11}{c}{Assignment} & Energy & Unc & NumTrans & u/g & Symmetry \\
	\hline
 0 & 0 & 0 & 0 & 0 & 0 & 0 & 0 &  1 & e & ortho & 2.35329 & 0.00003 & 204 & g & sigma\_g\_plus \\
 0 & 0 & 0 & 0 & 0 & 0 & 0 & 0 &  3 & e & ortho & 14.11952 & 0.00002 & 289 & g & sigma\_g\_plus \\
 0 & 0 & 0 & 0 & 0 & 0 & 0 & 0 &  5 & e & ortho & 35.29793 & 0.00002 & 306 & g & sigma\_g\_plus \\
 0 & 0 & 0 & 0 & 0 & 0 & 0 & 0 &  7 & e & ortho & 65.88710 & 0.00002 & 298 & g & sigma\_g\_plus \\
 0 & 0 & 0 & 0 & 0 & 0 & 0 & 0 &  9 & e & ortho & 105.88501 & 0.00002 & 306 & g & sigma\_g\_plus \\
 0 & 0 & 0 & 0 & 0 & 0 & 0 & 0 & 11 & e & ortho & 155.28899 & 0.00002 & 306 & g & sigma\_g\_plus \\
 0 & 0 & 0 & 0 & 0 & 0 & 0 & 0 & 13 & e & ortho & 214.09576 & 0.00002 & 306 & g & sigma\_g\_plus \\
 0 & 0 & 0 & 0 & 0 & 0 & 0 & 0 & 15 & e & ortho & 282.30144 & 0.00002 & 310 & g & sigma\_g\_plus \\
 0 & 0 & 0 & 0 & 0 & 0 & 0 & 0 & 17 & e & ortho & 359.90150 & 0.00002 & 294 & g & sigma\_g\_plus \\
 0 & 0 & 0 & 0 & 0 & 0 & 0 & 0 & 19 & e & ortho & 446.89078 & 0.00003 & 282 & g & sigma\_g\_plus \\
 0 & 0 & 0 & 0 & 0 & 0 & 0 & 0 & 21 & e & ortho & 543.26353 & 0.00002 & 274 & g & sigma\_g\_plus \\
 0 & 0 & 0 & 1 & 1 & 0 & 0 & 1 &  1 & e & ortho & 614.04436 & 0.00018 & 98 & g & pi\_g \\
 0 & 0 & 0 & 1 & 1 & 0 & 0 & 1 &  2 & f & ortho & 618.77696 & 0.00013 & 133 & g & pi\_g \\
	\hline
\end{tabular}
\end{table}

\begin{figure}[H]
	\includegraphics[width=0.8\linewidth]{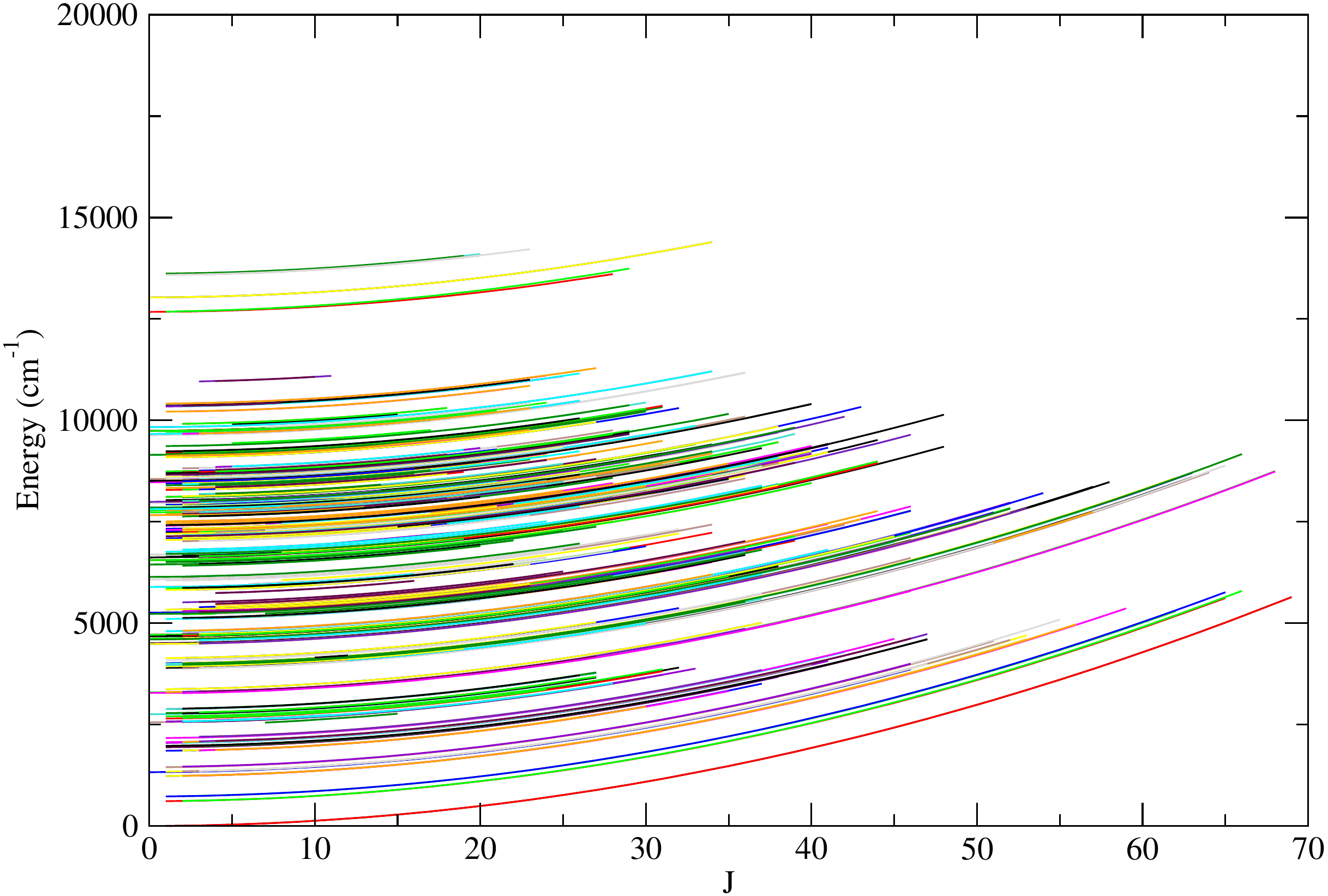}
	\caption{\Marvel\ energy levels (cm$^{-1}$) as a function of rotational quantum number, $J$,
for all the vibrational energy bands in the ortho network analysed in this
paper.}
	\label{fig:J_energy_ortho}
\end{figure}

\begin{figure}[H]
\includegraphics[width=0.8\linewidth]{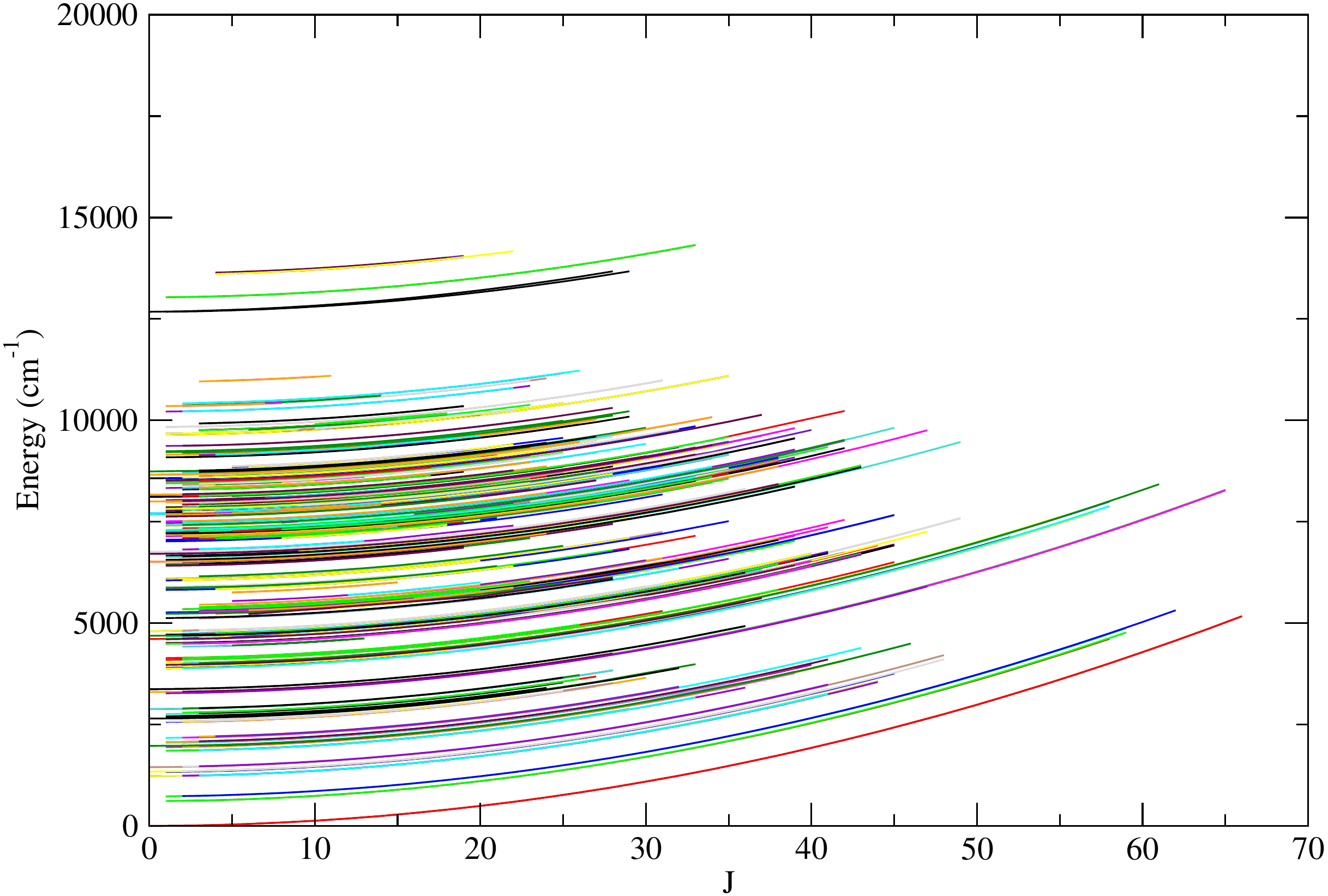}
	\caption{\Marvel\ energy levels (cm$^{-1}$) as a function of rotational quantum number, $J$,
for all the vibrational energy bands in the para network analysed in this
paper.}
	\label{fig:J_energy_para}
\end{figure}

\begin{figure}[H]
\includegraphics[width=\textwidth]{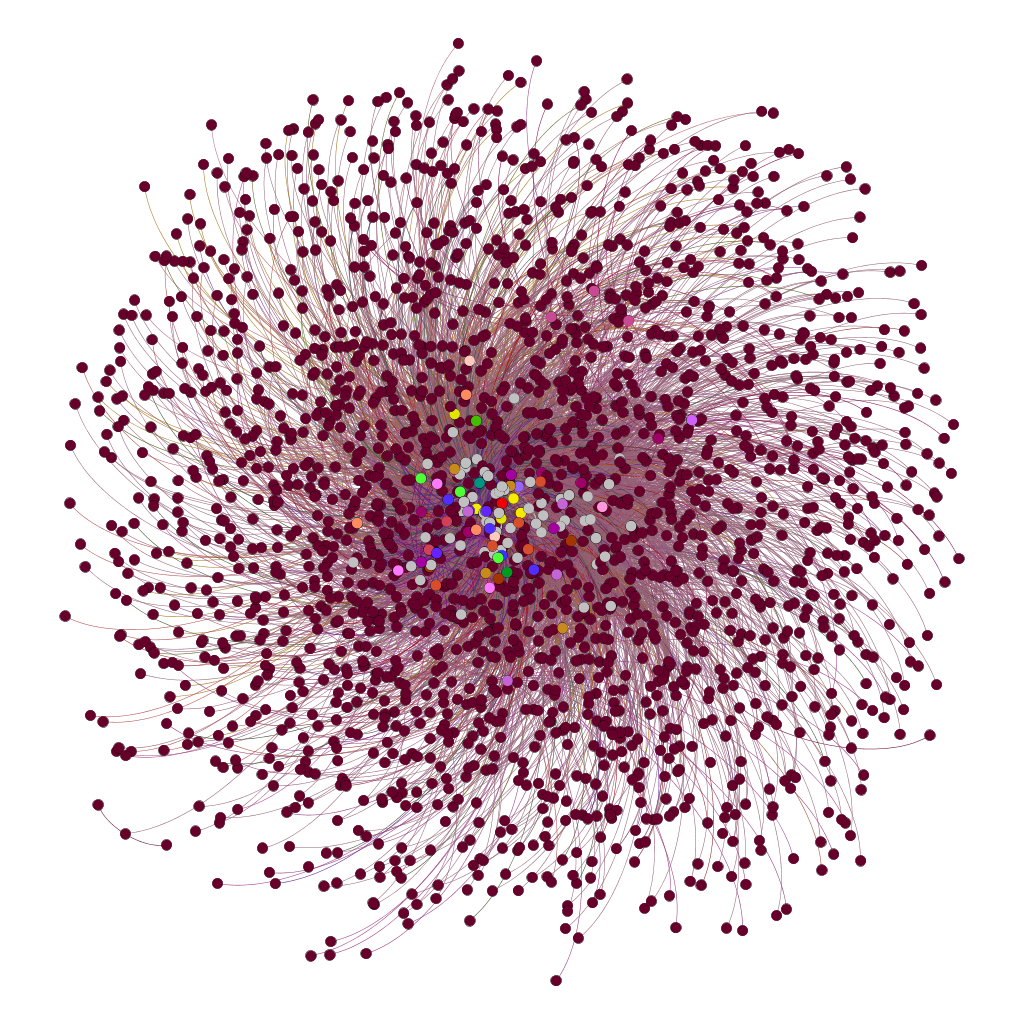}
	\caption{The \textit{ortho} component of the spectroscopic network produced using \Marvel\ input data.}
	\label{fig:ortho_spectro}
\end{figure}

\begin{figure}[H]
\includegraphics[width=\textwidth]{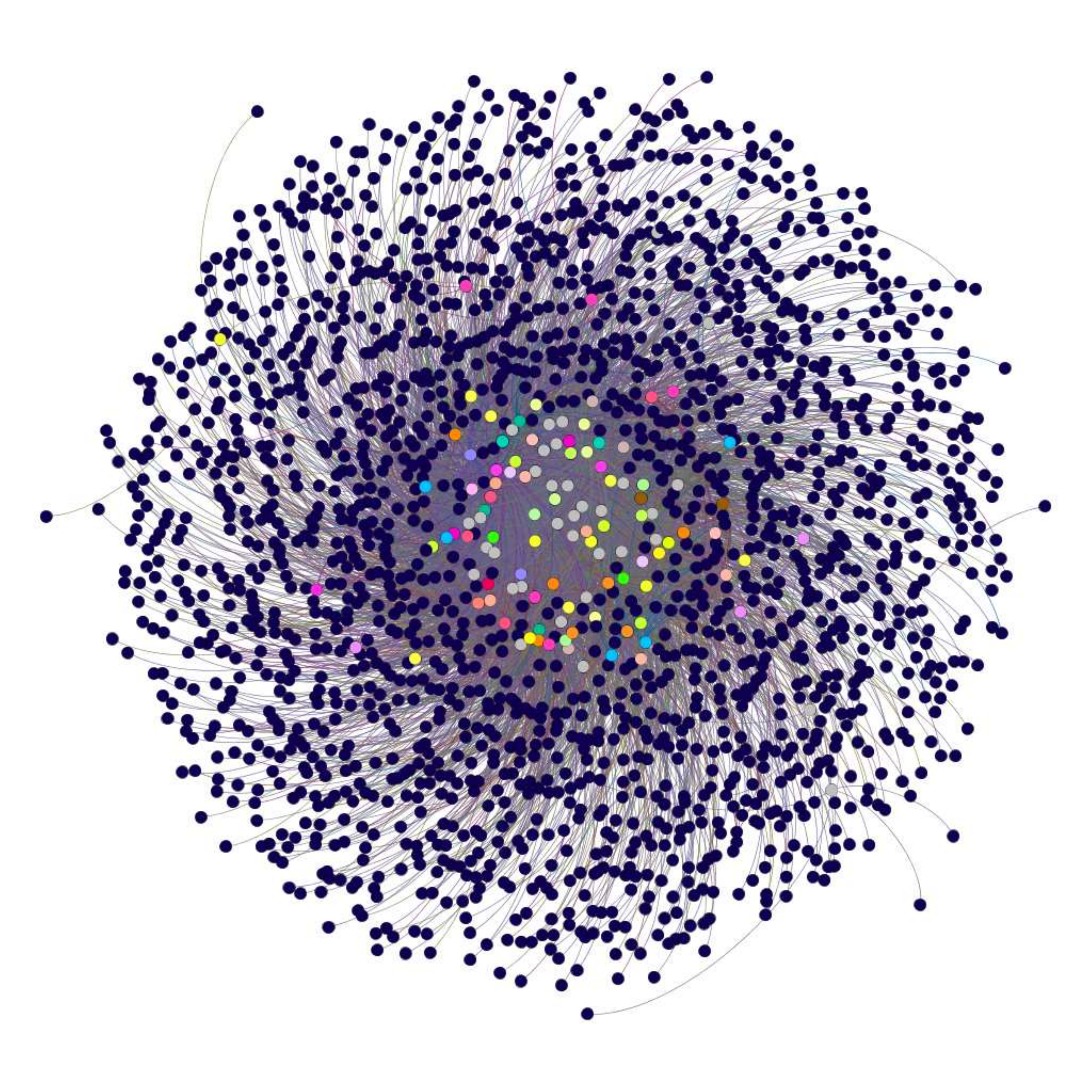}
	\caption{The \textit{para} component of the spectroscopic network produced using \Marvel\ input data.}
	\label{fig:para_spectro}
\end{figure}

\begin{figure}[H]
	\includegraphics[width=0.45\textwidth]{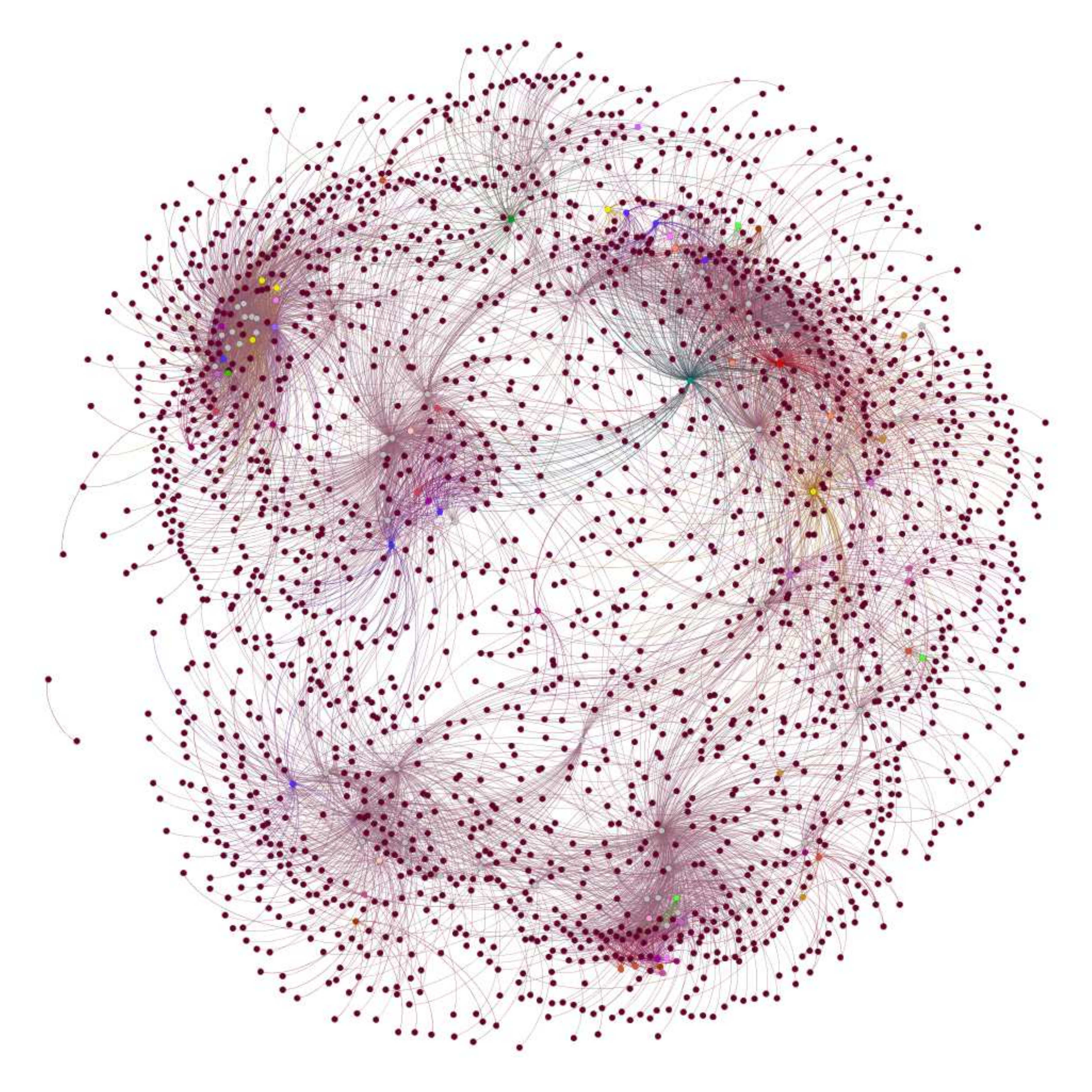}
	\includegraphics[width=0.55\textwidth]{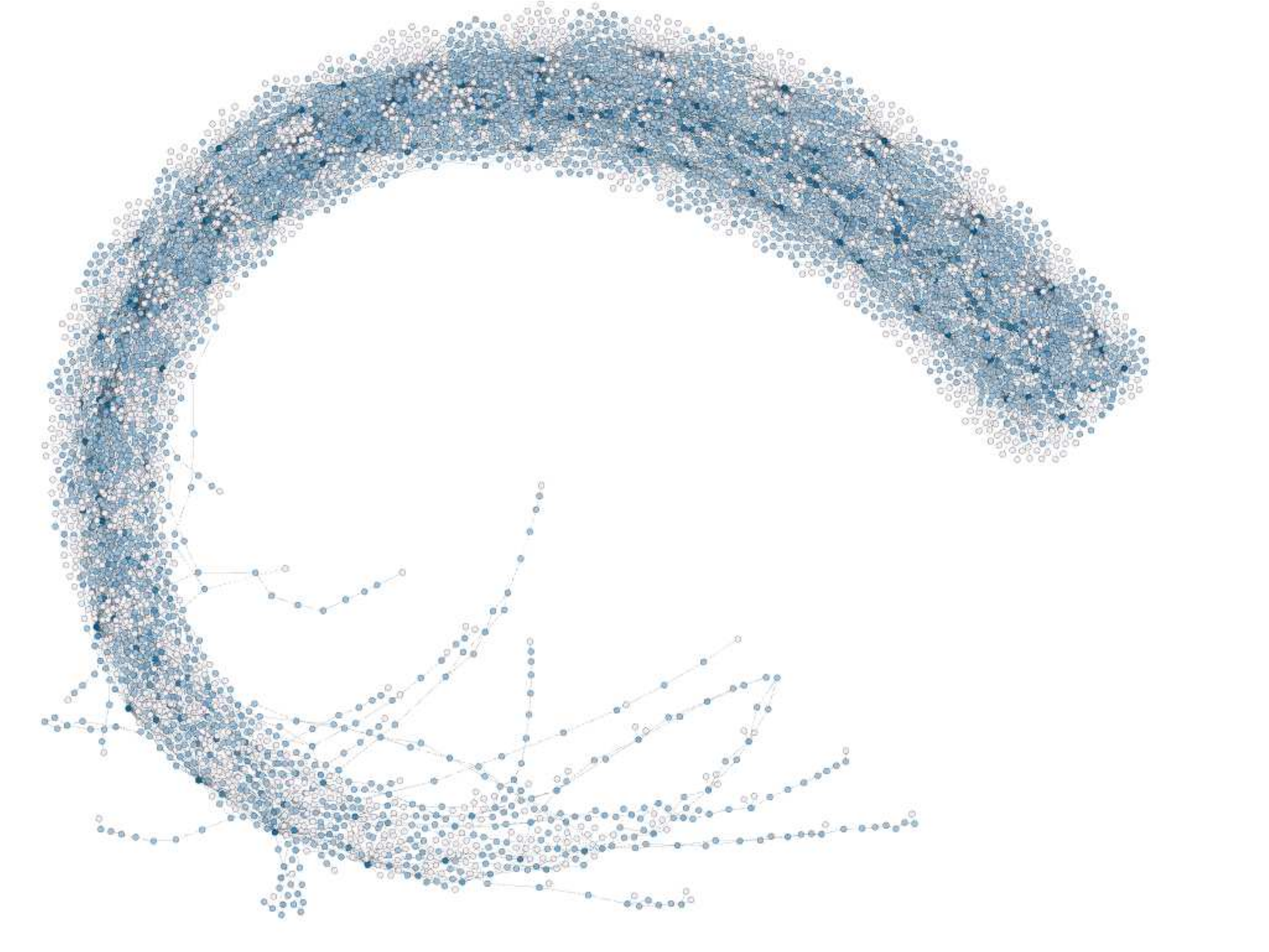}
	\caption{Alternative representations of the ortho (left) and para (right) spectroscopic networks produced using \Marvel\ input data.}
	\label{fig:orthopara_spectro}
\end{figure}

A total of 37,813 transitions were collated and considered (20,717 ortho and 17,096 para) from the data sources detailed in section 3. Of those 607 were found to be
inconsistent with others (353 ortho and 254 para) and thus removed from the final data set, leaving a total of 37,206 transitions used as input into \Marvel\ (20,364 ortho and 16,842 para).
A plot of energy as a function of rotational quantum number, $J$, was made
for each vibrational band as a check that quantum numbers had been assigned
consistently. Figure \ref{fig:J_energy_ortho} and \ref{fig:J_energy_para} show this for each vibrational band, for the
ortho and para states respectively.
Figures \ref{fig:ortho_spectro} and \ref{fig:para_spectro} illustrate the ortho and para spectroscopic networks, respectively. The nodes are energy levels and the edges the transitions between them. Each consists of a large main network with a series of smaller networks currently unattached. Different algorithms can be used to present these networks in a variety of ways; figure \ref{fig:orthopara_spectro}, for example, gives alternative representations of the structure. They highlight the intricate relationships between different energy levels and illustrate how the variety of sources collated in this work link together. We note that the inclusion of transitions intensities as weights in the spectroscopic network can aid in the determination of transitions which should preferentially be investigated in new experiments \cite{11CsFuxx.marvel}.


%

Table~\ref{t:vib} gives the vibrational ($J$=0) energies resulting from the \Marvel\ analysis, with associated uncertainity, vibrational assignment and the number of transitions (NumTrans) which were linked to the particular energy level. The higher the number of transitions the more certainty can be given to the energy value. See comment (3o) of section 3.1 relating to the band $(0122^{2}2^{-2})^{0}$ which may not have the correct assignment.

{\footnotesize
\begin{longtable}{rccccr}
	\caption{Vibrational energy levels (cm$^{-1}$) from \Marvel\ analysis
\label{t:vib}} \\
	\hline\hline
	$(v_1 v_2 v_3 v_4^{\ell_4} v_5^{\ell_5})^{K}$ & $e/f$ & State & \Marvel\ Energy (cm$^{-1}$) & Uncertainty (cm$^{-1}$) & NumTrans \\
	\hline
$(0000^{0}0^{0})^{0}$	&	e	&	para	&	0.000000	&	0.000000	&	85	\\
$(0002^{0}0^{0})^{0}$	&	e	&	para	&	1230.390303	&	0.000559	&	10	\\
$(0001^{1}1^{-1})^{0}$	&	e	&	ortho	&	1328.073466	&	0.000319	&	19	\\
$(0001^{1}1^{-1})^{0}$	&	f	&	para	&	1340.550679	&	0.001551	&	9	\\
$(0000^{0}2^{0})^{0}$	&	e	&	para	&	1449.112363	&	0.001189	&	10	\\
$(0100^{0}0^{0})^{0}$	&	e	&	para	&	1974.316617	&	0.006000	&	1	\\
$(0003^{1}1^{-1})^{0}$	&	e	&	ortho	&	2560.594937	&	0.002000	&	3	\\
$(0002^{2}2^{-2})^{0}$	&	e	&	para	&	2648.014468	&	0.004000	&	1	\\
$(0001^{1}3^{-1})^{0}$	&	e	&	ortho	&	2757.797907	&	0.001897	&	3	\\
$(0000^{0}4^{0})^{0}$	&	e	&	para	&	2880.220077	&	0.004000	&	1	\\
$(0101^{1}1^{-1})^{0}$	&	e	&	ortho	&	3281.899025	&	0.001744	&	5	\\
$(0010^{0}0^{0})^{0}$	&	e	&	ortho	&	3294.839579	&	0.001903	&	4	\\
$(0101^{1}1^{-1})^{0}$	&	f	&	para	&	3300.635590	&	0.007682	&	2	\\
$(1000^{0}0^{0})^{0}$	&	e	&	para	&	3372.838987	&	0.016000	&	1	\\
$(0103^{1}1^{-1})^{0}$	&	e	&	ortho	&	4488.838166	&	0.001200	&	2	\\
$(0012^{0}0^{0})^{0}$	&	e	&	ortho	&	4508.012219	&	0.002666	&	4	\\
$(0102^{2}2^{-2})^{0}$	&	f	&	ortho	&	4599.774669	&	0.003905	&	2	\\
$(0011^{1}1^{-1})^{0}$	&	e	&	para	&	4609.341046	&	0.005902	&	3	\\
$(0011^{1}1^{-1})^{0}$	&	f	&	ortho	&	4617.925870	&	0.005083	&	4	\\
$(1001^{1}1^{-1})^{0}$	&	e	&	ortho	&	4673.631058	&	0.001789	&	3	\\
$(1001^{1}1^{-1})^{0}$	&	f	&	para	&	4688.846488	&	0.011400	&	1	\\
$(0101^{1}3^{-1})^{0}$	&	e	&	ortho	&	4710.739822	&	0.018000	&	1	\\
$(0010^{0}2^{0})^{0}$	&	e	&	ortho	&	4727.069907	&	0.001193	&	3	\\
$(1000^{0}2^{0})^{0}$	&	e	&	para	&	4800.137287	&	0.000600	&	1	\\
$(0201^{1}1^{-1})^{0}$	&	e	&	ortho	&	5230.229286	&	0.010000	&	1	\\
$(0110^{0}0^{0})^{0}$	&	e	&	ortho	&	5260.021842	&	0.003328	&	2	\\
$(0103^{1}3^{-1})^{0}$	&	e	&	ortho	&	5893.260496	&	0.010000	&	1	\\
$(1001^{1}3^{-1})^{0}$	&	e	&	ortho	&	6079.693064	&	0.003714	&	2	\\
$(0010^{0}4^{0})^{0}$	&	e	&	ortho	&	6141.127536	&	0.010000	&	1	\\
$(0112^{0}0^{0})^{0}$	&	e	&	ortho	&	6449.106486	&	0.006000	&	1	\\
$(1102^{0}0^{0})^{0}$	&	e	&	para	&	6513.991447	&	0.008000	&	1	\\
$(1010^{0}0^{0})^{0}$	&	e	&	ortho	&	6556.464783	&	0.000100	&	4	\\
$(1101^{1}1^{-1})^{0}$	&	e	&	ortho	&	6623.139603	&	0.011915	&	2	\\
$(0110^{0}2^{0})^{0}$	&	e	&	ortho	&	6690.577636	&	0.012000	&	1	\\
$(2000^{0}0^{0})^{0}$	&	e	&	para	&	6709.021187	&	0.003714	&	2	\\
$(1100^{0}2^{0})^{0}$	&	e	&	para	&	6759.239077	&	0.010000	&	1	\\
$(0114^{0}0^{0})^{0}$	&	e	&	ortho	&	7665.441780	&	0.010000	&	1	\\
$(0022^{0}0^{0})^{0}$	&	e	&	para	&	7686.078947	&	0.002000	&	1	\\
$(0204^{2}2^{-2})^{0}$	&	e	&	para	&	7707.277687	&	0.004000	&	1	\\
$(1012^{0}0^{0})^{0}$	&	e	&	ortho	&	7732.793472	&	0.005291	&	4	\\
$(0203^{3}3^{-3})^{0}$	&	e	&	ortho	&	7787.324394	&	0.010000	&	1	\\
$(0021^{1}1^{-1})^{0}$	&	e	&	ortho	&	7805.004672	&	0.001876	&	3	\\
$(1103^{1}1^{-1})^{0}$	&	e	&	ortho	&	7816.006736	&	0.010000	&	1	\\
$(1011^{1}1^{-1})^{0}$	&	f	&	ortho	&	7853.277113	&	0.012000	&	1	\\
$(1010^{0}2^{0})^{0}$	&	e	&	ortho	&	7961.820133	&	0.007660	&	3	\\
$(2001^{1}1^{-1})^{0}$	&	e	&	ortho	&	7994.394918	&	0.002578	&	2	\\
$(2001^{1}1^{-1})^{0}$	&	f	&	para	&	8001.204086	&	0.009877	&	2	\\
$(2000^{0}2^{0})^{0}$	&	e	&	para	&	8114.362883	&	0.003705	&	3	\\
$(1100^{0}4^{0})^{0}$	&	e	&	para	&	8164.554028	&	0.008000	&	1	\\
$(1110^{0}0^{0})^{0}$	&	e	&	ortho	&	8512.056241	&	0.000429	&	3	\\
$(1201^{1}1^{-1})^{0}$	&	e	&	ortho	&	8556.589655	&	0.010000	&	1	\\
$(1201^{1}1^{-1})^{0}$	&	f	&	para	&	8570.322888	&	0.010000	&	1	\\
$(2100^{0}0^{0})^{0}$	&	e	&	para	&	8661.149087	&	0.010000	&	1	\\
$(0300^{0}4^{0})^{0}$	&	e	&	para	&	8739.814487	&	0.010000	&	1	\\
$(0310^{0}0^{0})^{0}$	&	e	&	ortho	&	9151.727686	&	0.010000	&	1	\\
$(0030^{0}0^{0})^{0}$	&	e	&	ortho	&	9639.863579	&	0.015435	&	2	\\
$(1112^{0}0^{0})^{0}$	&	e	&	ortho	&	9668.161468	&	0.015435	&	2	\\
$(0122^{2}2^{-2})^{0}$	&	f	&	ortho	&	9741.622286	&	0.030000	&	1	\\
$(0121^{1}1^{-1})^{0}$	&	e	&	ortho	&	9744.541486	&	0.030000	&	1	\\
$(2010^{0}0^{0})^{0}$	&	e	&	ortho	&	9835.173105	&	0.015435	&	2	\\
$(1030^{0}0^{0})^{0}$	&	e	&	ortho	&	12675.677286	&	0.001000	&	1	\\
$(3010^{0}0^{0})^{0}$	&	e	&	ortho	&	13033.293786	&	0.010000	&	1	\\
$(2210^{0}0^{0})^{0}$	&	e	&	ortho	&	13713.845686	&	0.006000	&	1	\\
	\hline\hline
\end{longtable}
}

\section{Comparison to other derived energy levels}

Table~\ref{t:compare} compares our rotational energy levels for the vibrational
ground state, which are determined up to $J=69$, with those obtained
by 16AmFaHe \cite{16AmFaHe.C2H2} from an effective Hamiltonian fit to
the observed data. In general the agreement is excellent. However for
the highest few levels with $J \geq 55$ we find differences which are
significantly larger than our uncertainties; our levels are
systematically below those of 16AmFaHe. This suggests that the
effective Hamiltonian treatment  used by 16AmFaHe becomes unreliable for these high $J$
levels. It should be noted that data relating to these highly excited levels originated from 16AmFaHe\_amy9, a high temperature experiment which has not been reproduced; see comment (3d), section 3.1. It is interesting to note that a further comparison with rotational energies extrapolated as part of 17LyPe's ASD-1000 spectroscopic databank \cite{17LyPe.C2H2}, also given in table~\ref{t:compare}, yields differences of approximately the same magnitude but, in contrast, consistently lower than our work.

{\footnotesize
\begin{longtable}{rccccccr}
	\caption{Comparison of pure rotational levels with
		those of 16AmFaHe \cite{16AmFaHe.C2H2}.
\label{t:compare}} \\
	\hline\hline
	$J$ & This work & Uncertainty & 16AmFaHe & Difference & 17LyPe & Difference & State \\
	\hline
1 & 2.35329 & 0.00003 & 2.353286417 & 0 & 2.3533 & 0.00001 & ortho \\
2 & 7.05982 & 0.00003 & 7.05982021 & 0 & 7.0598 & -0.00002 & para \\
3 & 14.11952 & 0.00002 & 14.119523294 & 0.00001 & 14.1195 & -0.00002 & ortho \\
4 & 23.53228 & 0.00003 & 23.532278547 & 0 & 23.5322 & -0.00008 & para \\
5 & 35.29793 & 0.00002 & 35.297929811 & 0 & 35.2978 & -0.00013 & ortho \\
6 & 49.41629 & 0.00003 & 49.416281896 & -0.00001 & 49.4161 & -0.00019 & para \\
7 & 65.88710 & 0.00002 & 65.887100587 & 0 & 65.8869 & -0.0002 & ortho \\
8 & 84.71012 & 0.00002 & 84.710112648 & -0.00001 & 84.7098 & -0.00032 & para \\
9 & 105.88501 & 0.00002 & 105.885005832 & 0 & 105.8846 & -0.00041 & ortho \\
10 & 129.41144 & 0.00003 & 129.411428888 & -0.00001 & 129.411 & -0.00044 & para \\
11 & 155.28899 & 0.00002 & 155.28899157 & 0.00001 & 155.2885 & -0.00049 & ortho \\
12 & 183.51727 & 0.00003 & 183.517264652 & -0.00001 & 183.5167 & -0.00057 & para \\
13 & 214.09576 & 0.00002 & 214.095779933 & 0.00002 & 214.0951 & -0.00066 & ortho \\
14 & 247.02403 & 0.00003 & 247.024030258 & 0 & 247.0233 & -0.00073 & para \\
15 & 282.30144 & 0.00002 & 282.301469525 & 0.00003 & 282.3007 & -0.00074 & ortho \\
16 & 319.92751 & 0.00003 & 319.927512702 & 0 & 319.9266 & -0.00091 & para \\
17 & 359.90150 & 0.00002 & 359.901535847 & 0.00004 & 359.9006 & -0.0009 & ortho \\
18 & 402.22287 & 0.00003 & 402.22287612 & 0.00001 & 402.2219 & -0.00097 & para \\
19 & 446.89078 & 0.00003 & 446.890831804 & 0.00006 & 446.8898 & -0.00098 & ortho \\
20 & 493.90464 & 0.00003 & 493.904662324 & 0.00002 & 493.9036 & -0.00104 & para \\
21 & 543.26353 & 0.00002 & 543.263588267 & 0.00006 & 543.2625 & -0.00103 & ortho \\
22 & 594.96668 & 0.00004 & 594.966791406 & 0.00011 & 594.9657 & -0.00098 & para \\
23 & 649.01328 & 0.00003 & 649.013414717 & 0.00014 & 649.0123 & -0.00098 & ortho \\
24 & 705.40237 & 0.00004 & 705.402562408 & 0.00019 & 705.4015 & -0.00087 & para \\
25 & 764.13315 & 0.00003 & 764.133299944 & 0.00015 & 764.1322 & -0.00095 & ortho \\
26 & 825.20439 & 0.00004 & 825.204654067 & 0.00026 & 825.2037 & -0.00069 & para \\
27 & 888.61531 & 0.00003 & 888.615612828 & 0.00031 & 888.6147 & -0.00061 & ortho \\
28 & 954.36496 & 0.00005 & 954.365125617 & 0.00017 & 954.3642 & -0.00076 & para \\
29 & 1022.45167 & 0.00003 & 1022.452103183 & 0.00044 & 1022.4513 & -0.00037 & ortho \\
30 & 1092.87513 & 0.00005 & 1092.875417676 & 0.00029 & 1092.8747 & -0.00043 & para \\
31 & 1165.63343 & 0.00004 & 1165.633902667 & 0.00048 & 1165.6333 & -0.00013 & ortho \\
32 & 1240.72592 & 0.00017 & 1240.726353188 & 0.00043 & 1240.7259 & -0.00002 & para \\
33 & 1318.15099 & 0.00011 & 1318.151525765 & 0.00054 & 1318.1512 & 0.00021 & ortho \\
34 & 1397.90769 & 0.00023 & 1397.908138445 & 0.00045 & 1397.908 & 0.00031 & para \\
35 & 1479.99435 & 0.00007 & 1479.994870843 & 0.00053 & 1479.9949 & 0.00055 & ortho \\
36 & 1564.40979 & 0.00026 & 1564.410364167 & 0.00057 & 1564.4105 & 0.00071 & para \\
37 & 1651.15189 & 0.00017 & 1651.153221265 & 0.00134 & 1651.1535 & 0.00161 & ortho \\
38 & 1740.22038 & 0.00037 & 1740.222006657 & 0.00163 & 1740.2225 & 0.00212 & para \\
39 & 1831.61393 & 0.00026 & 1831.615246582 & 0.00132 & 1831.6159 & 0.00197 & ortho \\
40 & 1925.33058 & 0.00074 & 1925.331429031 & 0.00085 & 1925.3322 & 0.00162 & para \\
41 & 2021.36757 & 0.00043 & 2021.369003793 & 0.00144 & 2021.3699 & 0.00233 & ortho \\
42 & 2119.72430 & 0.00057 & 2220.401938666 & 0.00134 & 2220.4029 & 0.0023 & ortho \\
44 & 2323.39201 & 0.00127 & 2323.394007739 & 0.002 & 2323.395 & 0.00299 & para \\
45 & 2428.69912 & 0.00135 & 2428.70088714 & 0.00177 & 2428.7018 & 0.00268 & ortho \\
46 & 2536.31702 & 0.00103 & 2536.320836316 & 0.00382 & 2536.3217 & 0.00468 & para \\
47 & 2646.25026 & 0.00128 & 2646.252076785 & 0.00182 & 2646.2527 & 0.00244 & ortho \\
48 & 2758.49217 & 0.00142 & 2758.492792187 & 0.00062 & 2758.4931 & 0.00093 & para \\
49 & 2873.03874 & 0.00194 & 2873.041128336 & 0.00239 & 2873.0411 & 0.00236 & ortho \\
50 & 2989.89046 & 0.00175 & 2989.895193269 & 0.00473 & 2989.8947 & 0.00424 & para \\
51 & 3109.04649 & 0.00148 & 3109.0530573 & 0.00657 & 3109.0519 & 0.00541 & ortho \\
52 & 3230.50478 & 0.00124 & 3230.512753073 & 0.00797 & 3230.5108 & 0.00602 & para \\
53 & 3354.26378 & 0.00224 & 3354.272275619 & 0.0085 & 3354.2694 & 0.00562 & ortho \\
54 & 3480.32661 & 0.00250 & 3480.329582411 & 0.00297 & 3480.3256 & -0.00101 & para \\
55 & 3608.67187 & 0.00250 & 3608.682593419 & 0.01073 & 3608.6772 & 0.00533 & ortho \\
56 & 3739.32523 & 0.00118 & 3739.329191172 & 0.00396 & 3739.3223 & -0.00293 & para \\
57 & 3872.25530 & 0.00208 & 3872.267220814 & 0.01193 & 3872.2585 & 0.0032 & ortho \\
58 & 4007.49264 & 0.00170 & 4007.494490165 & 0.00185 & 4007.4836 & -0.00904 & para \\
59 & 4144.99542 & 0.00118 & 4145.008769784 & 0.01335 & 4144.9955 & 0.00008 & ortho \\
60 & 4284.80143 & 0.00181 & 4284.807793029 & 0.00636 & 4284.7918 & -0.00963 & para \\
61 & 4426.87720 & 0.00154 & 4426.889256124 & 0.01206 & 4426.8701 & -0.0071 & ortho \\
62 & 4571.24409 & 0.00142 & 4571.25081822 & 0.00673 & 4571.2281 & -0.01599 & para \\
63 & 4717.87442 & 0.00142 & 4717.890101462 & 0.01569 & 4717.8635 & -0.01092 & ortho \\
64 & 4866.79028 & 0.00232 & 4866.804691055 & 0.01441 & 4866.7736 & -0.01668 & para \\
65 & 5017.97095 & 0.00168 & 5017.992135336 & 0.02119 & 5017.9561 & -0.01485 & ortho \\
66 & 5171.43923 & 0.00366 & 5171.449945837 & 0.01072 & 5171.4085 & -0.03073 & para \\
67 & 5327.14526 & 0.00195 & 5327.175597358 & 0.03034 & 5327.128 & -0.01726 & ortho \\
69 & 5645.38676 & 0.00300 & 5645.420139428 & 0.03338 & 5645.3585 & -0.02826 & ortho\\
	\hline\hline
\end{longtable}
}

\begin{figure}
	\includegraphics[width=0.8\linewidth]{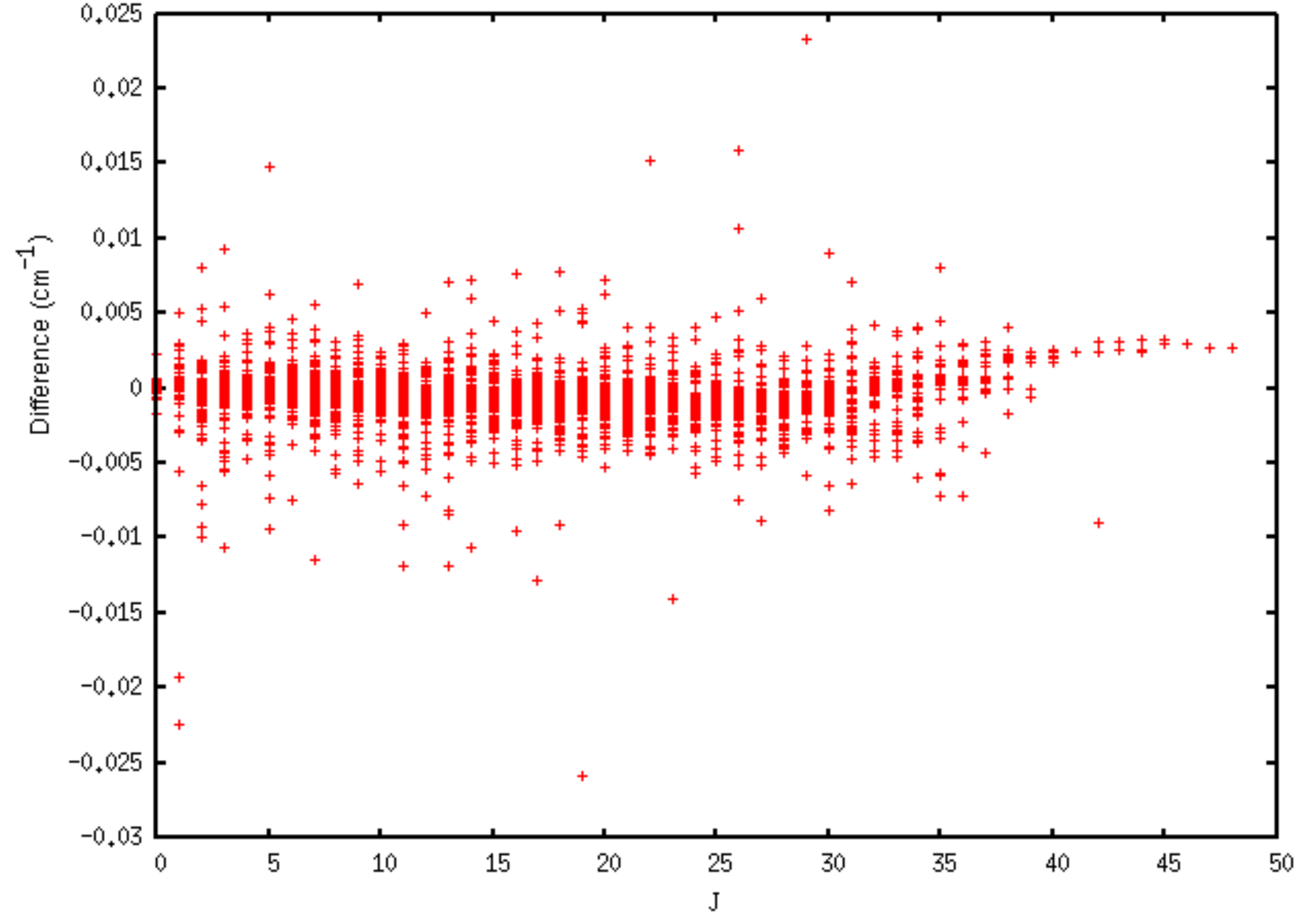}
	\caption{Differences between the energy term values given in 17LyCa \cite{17LyCaxx.C2H2} and this work as a function of rotational angular momentum quantum number, $J$.}
	\label{fig:J_comp_17LyCa}
\end{figure}

The supplementary data from 17LyCa \cite{17LyCaxx.C2H2} contains lower
energy levels, frequency and assignments, from which upper energy levels can be
calculated. Figure \ref{fig:J_comp_17LyCa} gives the differences between the energies
given in 17LyCa and this work as a function of $J$. The vast majority
are within 0.005~cm$^{-1}$. Note that the difference in labelling of
some bands has been taken into account when comparisons are made (see
comment (3l) in section 3.1 and comment (1b) in section 3.2).

\begin{figure}[H]
	\includegraphics[width=\linewidth]{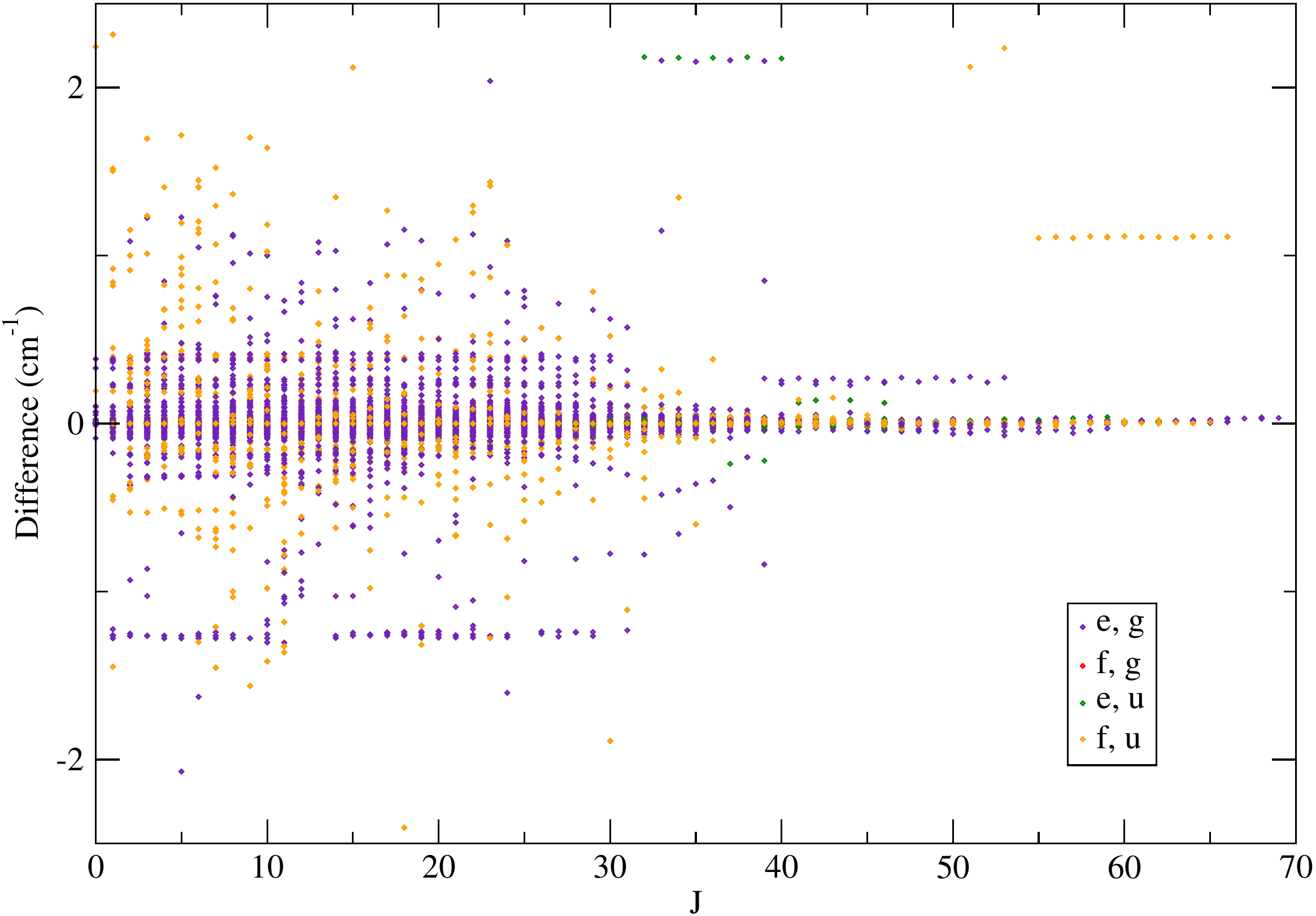}
	\caption{Deviations in cm$^{-1}$ between this work and 16AmFaHe \cite{16AmFaHe.C2H2} as a function of rotational angular momentum quantum number, $J$. Different colours represent different designations of $e/f$ and $u/g$.}
	\label{fig:J_16AmFaHe}
\end{figure}

The energy levels given as supplementary data in annex 5 of 16AmFaHe
\cite{16AmFaHe.C2H2} are separated into polyads which are characterised by a
small number of quantum numbers; $N_{rm v}=5v_1+3v_2+5v_3+v_4+v_5$, $J$, $e/f$
symmetry and $u/g$ symmetry. As there are more than one state defined by these
quantum numbers, the only comparison that was possible to make was to match
these and find the closest energy value within these bounds. As such, we cannot
be certain that bands have been matched correctly. 17LyCa compared what they
could against 16AmFaHe's data but also could not find a reliable way to
determine unambiguously which energy of each polyad block corresponds to their
energy levels. Figure \ref{fig:J_16AmFaHe} gives the difference between the
energies in this work and those matched with 16AmFaHe as a function of
rotational angular momentum quantum number, $J$. 
6160 out of the 11,154 energies differ by less than 0.01~cm$^{-1}$. However, this
leaves 4994 energies with a difference of higher than 0.01~cm$^{-1}$. 2176 of these
energies also appear in 17LyCa, so a comparison could be made between the three.
Only 7 of the energies in the 17LyCa dataset are closer to 16AmFaHe than this
work, and of those all are within 0.02~cm$^{-1}$ with this work.

\begin{figure}[H]
	\includegraphics[width=\linewidth]{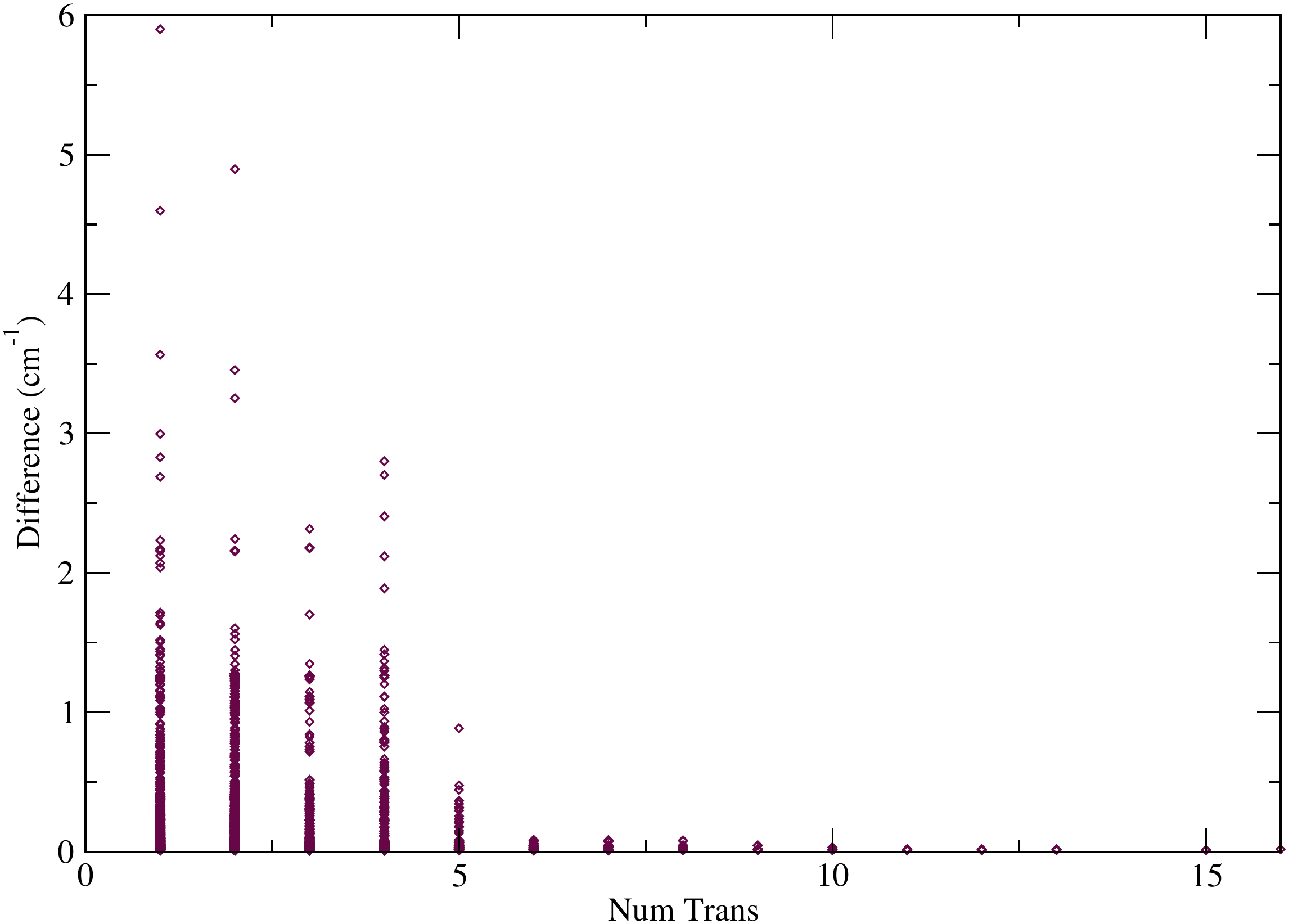}
	\caption{Deviations in cm$^{-1}$ between this work and 16AmFaHe \cite{16AmFaHe.C2H2} as a function of the number of transitions that link to the energy level in our dataset.}
	\label{fig:numtrans_16Am}
\end{figure}

It should be noted, however, that the differences between this work and 16AmFaHe are largest for those energy levels with a low value of NumTrans (the number of transitions that link the energy state to other energies within the dataset); see figure \ref{fig:numtrans_16Am}. The vast majority of energy levels which only have one transition are not in the 17LyCa dataset. Many of these transitions came from the data source 16AmFaHe\_amy09; see comment (3d) in section 3.1. It would be of use to have more experimental data on transitions to these levels in order to confirm their validity. The entire band $(0122^22^{-2})^0$ has differences of over 900cm-1 in comparison to the matched values in 16AmFaHe. This indicates that this band has been misassigned (see comment (3o in section 3.1)). We are uncertain currently as what it should be reassigned to. We have excluded this band from figures \ref{fig:J_16AmFaHe} and \ref{fig:numtrans_16Am}.

It should be made clear, as mentioned above, that those energy levels present in the input data which are only linked to the main components of the spectroscopic network by one transition should be treated with caution; this number is given as a parameter in the third to last column of the output files included in the supplementary data. It can be used, along with the uncertainties, as an indication of the reliability of each energy level. Note, finally, that \Marvel\ only processes data given as input; it does not extrapolate to higher excitations.

\section{Conclusions}

A total of 37,813 measured experimental transitions from 61 publications have been considered in this work. From this 6013 ortho and 5200 para energy levels have been determined using the Measured Active Rotational-Vibrational Energy Levels (\Marvel) technique.
These results have been compared with alternative compilations
based on the use of effective Hamiltonians. An \textit{ab initio} high temperature linelist for acetylene is in preparation as part of the ExoMol project \cite{jtlinTROVE}, for which this data will be used in the process of validation of theoretical calculations.

A significant part of this work was performed by  pupils from Highams Park
School in London, as part of
a project known as ORBYTS (Original Research By Young Twinkle Scientists).
The \Marvel\ study of TiO \cite{jt672} was also performed as part of the ORBYTS project
and further studies on other key molecules will be published in due course.
A paper discussing our experiences of performing original research in collaboration
with school children will be published elsewhere \cite{jt709}.

\section{Supplementary Data}\label{sec:fin}

Supplementary material associated with this article can be found, in the online version, at 10.1016/j.jqsrt.2017.08.018. There are four files provided, as listed in Table~\ref{t:SI}. The column definitions are given in Table~\ref{t:def} for files 1 and 2 (\Marvel\ input files) and Table~\ref{t:def34} for files 3 and 4 (\Marvel\ output files).

\begin{table}[H]
	\caption{Supplementary data files.}
\label{t:SI}
\tt
	\begin{tabular}{cc}
		\hline\hline
		File & Name\\
		\hline
		1 & MARVEL\_ortho\_transitions\_input.txt \\
		2 & MARVEL\_para\_transitions\_input.txt\\
		3 & MARVELenergylevels\_ortho\_output.txt\\
		4 & MARVELenergylevels\_para\_output.txt \\
		\hline\hline
	\end{tabular}
\end{table}

\begin{table}[H]
	\caption{Definition of columns in files 1 and 2.}
\label{t:def}
\footnotesize
	\begin{tabular}{rcc}
		\hline\hline
		Column & Label & Description\\
		\hline
		1 & Energy (cm$^{-1}$)	& Transition wavenumber \\
		2 & Uncertainty (cm$^{-1}$)	& Associated uncertainity \\
		& Upper assignments: & \\
		3 & $v_1$	& CH symmetric stretch ($\sigma_g^{+}$) \\
		4 & $v_2$	& CC symmetric stretch ($\sigma_g^{+}$) \\
		5 & $v_3$	& CH antisymmetric stretch ($\sigma_u^{+}$) \\
		6 & $v_4$	& Symmetric (trans) bend ($\pi_g$) \\
		7 & $\ell_4$	& Vibrational angular momentum associated with $v_4$ \\
		8 & $v_5$	& Antisymmetric (cis) bend ($\pi_u$) \\
		9 & $\ell_5$	& Vibrational angular momentum associated with $v_5$ \\
		10 & $K$	& =$|\ell_4+\ell_5|$, total vibrational angular momentum \\
		11 & $J$	& Rotational angular momentum \\
		12 & $e/f$	& Symmetry relative to the Wang transformation (see section 2.2) \\
		13 & ortho/para	& Nuclear spin state (see section 2.2) \\
		& Lower assignments: & \\
		14 & $v_1$	& CH symmetric stretch ($\sigma_g^{+}$) \\
		15 & $v_2$	& CC symmetric stretch ($\sigma_g^{+}$) \\
		16 & $v_3$	& CH antisymmetric stretch ($\sigma_u^{+}$) \\
		17 & $v_4$	& Symmetric (trans) bend ($\pi_g$) \\
		18 & $\ell_4$	& Vibrational angular momentum associated with $v_4$ \\
		19 & $v_5$	& Antisymmetric (cis) bend ($\pi_u$) \\
		20 & $\ell_5$	& Vibrational angular momentum associated with $v_5$ \\
		21 & $K$	& =$|\ell_4+\ell_5|$, total vibrational angular momentum \\
		22 & $J$	& Rotational angular momentum \\
		23 & $e/f$	& Symmetry relative to the Wang transformation (see section 2.2) \\
		24 & ortho/para	& Nuclear spin state (see section 2.2) \\
		25 & Ref & Unique reference label (see section 2.2) \\
		\hline\hline
	\end{tabular}
\end{table}

\begin{table}[H]
	\caption{Definition of columns in files 3 and 4.}
\label{t:def34}
\footnotesize
	\begin{tabular}{rcc}
		\hline\hline
		Column & Label & Description\\
		\hline
		1 & $v_1$	& CH symmetric stretch ($\sigma_g^{+}$) \\
		2 & $v_2$	& CC symmetric stretch ($\sigma_g^{+}$) \\
		3 & $v_3$	& CH antisymmetric stretch ($\sigma_u^{+}$) \\
		4 & $v_4$	& Symmetric (trans) bend ($\pi_g$) \\
		5 & $\ell_4$	& Vibrational angular momentum associated with $v_4$ \\
		6 & $v_5$	& Antisymmetric (cis) bend ($\pi_u$) \\
		7 & $\ell_5$	& Vibrational angular momentum associated with $v_5$ \\
		8 & $K$	& =$|\ell_4+\ell_5|$, total vibrational angular momentum \\
		9 & $J$	& Rotational angular momentum \\
		10 & $e/f$	& Symmetry relative to the Wang transformation (see section 2.2) \\
		11 & ortho/para	& Nuclear spin state (see section 2.2) \\
		12 & Energy (cm$^{-1}$)	& \Marvel\ energy assignment \\
		13 & Uncertainty (cm$^{-1}$)	& \Marvel\ uncertainty \\
		14 & NumTrans	& The number of transitions in the dataset which link to this state \\
		15 & $u/g$ symmetry	& See section 2.2 \\
		16 & Symmetry label	& See section 2.2 \\
		\hline\hline
	\end{tabular}
\end{table}

\section{Acknowledgements}

We would like to thank Jon Barker, Fawad Sheikh, and Sheila Smith from Highams Park School for continued support and enthusiasm. Jean Vander Auwera, Juho Karhu, Alain Campargue, Oleg Lyulin and Michel Herman are thanked for helping with our queries and
providing digital versions of published data where necessary. We are grateful to Laura McKemmish for helpful discussions
and advice. We are also grateful to the authors of all the experimental sources analysed in this work
(and apologies to any we may have missed) for the high accuracy and copious
amounts of acetylene data they have made available to the scientific community over the years.
This work was supported by STFC Project ST/J002925, by NKFIH by grant no. 119658, the ERC under Advanced Investigator Project 267219
and the COST action MOLIM: Molecules in Motion (CM1405). We would like to thank the reviewers of this paper for useful comments and suggestions, and to Michel Herman, Badr Amyay, Oleg Lyulin, and Alain Campargue for interesting discussions. 

\bibliographystyle{elsarticle-num}
\bibliography{journals_phys,jtj,C2H2,MARVEL,HCNO,linear,exogen,dwarfs,exoplanets,methods,linelists}

\begin{thebibliography}{100}
\expandafter\ifx\csname url\endcsname\relax
  \def\url#1{\texttt{#1}}\fi
\expandafter\ifx\csname urlprefix\endcsname\relax\def\urlprefix{URL }\fi
\expandafter\ifx\csname href\endcsname\relax
  \def\href#1#2{#2} \def\path#1{#1}\fi

\bibitem{12Gaxxxx.C2H2}
A.~Gaydon, The spectroscopy of flames, Springer Science \& Business Media,
  2012.

\bibitem{10MeScSk.C2H2}
M.~Mets{\"a}l{\"a}, F.~M. Schmidt, M.~Skytta, O.~Vaittinen, L.~Halonen,
  {Acetylene in breath: background levels and real-time elimination kinetics
  after smoking}, J. Breath Res. 4 (2010) 046003.
\newblock \href {http://dx.doi.org/10.1088/1752-7155/4/4/046003}
  {\path{doi:10.1088/1752-7155/4/4/046003}}.

\bibitem{14DhRaxx.C2H2}
H.~Dhanoa, J.~M.~C. Rawlings, {Is acetylene essential for carbon dust
  formation?}, Mon. Not. R. Astron. Soc. 440 (2014) 1786--1793.
\newblock \href {http://dx.doi.org/10.1093/mnras/stu401}
  {\path{doi:10.1093/mnras/stu401}}.

\bibitem{76RiHaKl.C2H2}
S.~T. Ridgway, D.~N.~B. Hall, S.~G. Kleinmann, D.~A. Weinberger, R.~S. Wojslaw,
  {Circumstellar acetylene in the infrared spectrum of IRC+ 10� 216}, Nature
  264 (1976) 345--346.

\bibitem{13BiRiHe.dwarfs}
C.~Bilger, P.~Rimmer, C.~Helling, Small hydrocarbon molecules in cloud-forming
  brown dwarf and giant gas planet atmospheres, Mon. Not. R. Astron. Soc. 435
  (2013) 1888--1903.
\newblock \href {http://dx.doi.org/10.1093/mnras/stt1378}
  {\path{doi:10.1093/mnras/stt1378}}.

\bibitem{82RiBaRa.C2H2}
C.~P. Rinsland, A.~Baldacci, K.~N. Rao, {Acetylene bands observed in carbon
  stars - a laboratory study and an illustrative example of its application to
  IRC+10216}, Astrophys. J. Suppl. 49 (1982) 487--513.
\newblock \href {http://dx.doi.org/10.1086/190808} {\path{doi:10.1086/190808}}.

\bibitem{04GaHoJo}
R.~Gautschy-Loidl, S.~Hofner, U.~Jorgensen, J.~Hron, {Dynamic model atmospheres
  of AGB stars - IV. A comparison of synthetic carbon star spectra with
  observations}, Astron. Astrophys. 422 (2004) 289--306.
\newblock \href {http://dx.doi.org/10.1051/0004-6361:20035860}
  {\path{doi:10.1051/0004-6361:20035860}}.

\bibitem{08OrVoxx.C2H2}
R.~S. Oremland, M.~A. Voytek, {Acetylene as fast food: Implications for
  development of life on anoxic primordial earth and in the outer solar
  system}, Astrobiology 8 (2008) 45--58.
\newblock \href {http://dx.doi.org/10.1089/ast.2007.0183}
  {\path{doi:10.1089/ast.2007.0183}}.

\bibitem{96BrToWe.C2H2}
T.~Y. Brooke, A.~T. Tokunaga, H.~A. Weaver, J.~Crovisier, D.~BockeleeMorvan,
  D.~Crisp, {Detection of acetylene in the infrared spectrum of comet
  Hyakutake}, Nature 383 (1996) 606--608.
\newblock \href {http://dx.doi.org/10.1038/383606a0}
  {\path{doi:10.1038/383606a0}}.

\bibitem{jt629}
A.~Tsiaras, M.~Rocchetto, I.~P. Waldmann, G.~Tinetti, R.~Varley, G.~Morello,
  E.~J. Barton, S.~N. Yurchenko, J.~Tennyson, {Detection of an atmosphere
  around the super-Earth 55 Cancri e}, Astrophys. J. 820 (2016) 99.
\newblock \href {http://dx.doi.org/10.3847/0004-637X/820/2/99}
  {\path{doi:10.3847/0004-637X/820/2/99}}.

\bibitem{07Hexxxx.C2H2}
M.~Herman, {The acetylene ground state saga}, Mol. Phys. 105 (2007) 2217--2241.
\newblock \href {http://dx.doi.org/10.1080/00268970701518103}
  {\path{doi:10.1080/00268970701518103}}.

\bibitem{10DiHexx.C2H2}
K.~Didriche, M.~Herman, {A four-atom molecule at the forefront of spectroscopy,
  intramolecular dynamics and astrochemistry: Acetylene}, Chem. Phys. Lett. 496
  (2010) 1--7.
\newblock \href {http://dx.doi.org/10.1016/j.cplett.2010.07.031}
  {\path{doi:10.1016/j.cplett.2010.07.031}}.

\bibitem{11Herman.C2H2}
M.~Herman, {High-resolution Infrared Spectroscopy of Acetylene: Theoretical
  Background and Research Trends}, John Wiley \& Sons, Ltd, 2011, pp.
  1993--2026.
\newblock \href {http://dx.doi.org/10.1002/9780470749593.hrs101}
  {\path{doi:10.1002/9780470749593.hrs101}}.

\bibitem{93BrHaxx.method}
M.~J. Bramley, N.~C. Handy, {Efficient calculation of rovibrational eigenstates
  of sequentially bonded 4-atom molecules}, J. Chem. Phys. 98 (1993)
  1378--1397.
\newblock \href {http://dx.doi.org/10.1063/1.464305}
  {\path{doi:10.1063/1.464305}}.

\bibitem{96Scxxxx.linear}
D.~W. Schwenke, {Variational calculations of rovibrational energy levels and
  transition intensities for tetratomic molecules}, J. Phys. Chem. 100 (1996)
  2867--2884.
\newblock \href {http://dx.doi.org/10.1021/jp9525447}
  {\path{doi:10.1021/jp9525447}}.

\bibitem{jt346}
I.~N. Kozin, M.~M. Law, J.~Tennyson, J.~M. Hutson, {Calculating energy levels
  of isomerizing tetraatomic molecules: II. The vibrational states of acetylene
  and vinylidene}, J. Chem. Phys. 122 (2005) 064309.

\bibitem{02XuLiXi.C2H2}
D.~G. Xu, G.~H. Li, D.~Q. Xie, H.~Guo, {Full-dimensional quantum calculations
  of vibrational energy levels of acetylene (HCCH) up to 13,000 cm$^{-1}$},
  Chem. Phys. Lett. 365 (2002) 480--486.
\newblock \href {http://dx.doi.org/10.1016/S0009-2614(02)01503-8}
  {\path{doi:10.1016/S0009-2614(02)01503-8}}.

\bibitem{03XuGuZo.C2H2}
D.~G. Xu, H.~Guo, S.~L. Zou, J.~M. Bowman, {A scaled ab initio potential energy
  surface for acetylene and vinylidene}, Chem. Phys. Lett. 377 (2003) 582--588.
\newblock \href {http://dx.doi.org/10.1016/S0009-2614(03)01184-9}
  {\path{doi:10.1016/S0009-2614(03)01184-9}}.

\bibitem{jt479}
A.~Urru, I.~N. Kozin, G.~Mulas, B.~J. Braams, J.~Tennyson, {Ro-vibrational
  spectra of C$_2$H$_2$ based on variational nuclear motion calcalculations},
  Mol. Phys. 108 (2010) 1973--1990.

\bibitem{jt528}
J.~Tennyson, S.~N. Yurchenko, {ExoMol: molecular line lists for exoplanet and
  other atmospheres}, Mon. Not. R. Astron. Soc. 425 (2012) 21--33.
\newblock \href {http://dx.doi.org/10.1111/j.1365-2966.2012.21440.x}
  {\path{doi:10.1111/j.1365-2966.2012.21440.x}}.

\bibitem{jt631}
J.~Tennyson, S.~N. Yurchenko, A.~F. Al-Refaie, E.~J. Barton, K.~L. Chubb, P.~A.
  Coles, S.~Diamantopoulou, M.~N. Gorman, C.~Hill, A.~Z. Lam, L.~Lodi, L.~K.
  McKemmish, Y.~Na, A.~Owens, O.~L. Polyansky, T.~Rivlin, C.~Sousa-Silva, D.~S.
  Underwood, A.~Yachmenev, E.~Zak, {The ExoMol database: molecular line lists
  for exoplanet and other hot atmospheres}, J. Mol. Spectrosc. 327 (2016)
  73--94.
\newblock \href {http://dx.doi.org/10.1016/j.jms.2016.05.002}
  {\path{doi:10.1016/j.jms.2016.05.002}}.

\bibitem{16AmFaHe.C2H2}
B.~Amyay, A.~Fayt, M.~Herman, J.~{Vander Auwera}, {Vibration-rotation
  spectroscopic database on acetylene, $X^1\Sigma_g^+$ $^{12}$C$_2$H$_2$}, J.
  Phys. Chem. Ref. Data 45 (2016) 023103.
\newblock \href {http://dx.doi.org/10.1063/1.4947297}
  {\path{doi:10.1063/1.4947297}}.

\bibitem{17LyCaxx.C2H2}
O.~M. Lyulin, A.~Campargue, {An empirical spectroscopic database for acetylene
  in the regions of 5850.6341 cm$^{-1}$ and 7000.9415 cm$^{-1}$}, J. Quant.
  Spectrosc. Radiat. Transf. (2017) --\href
  {http://dx.doi.org/10.1016/j.jqsrt.2017.01.036}
  {\path{doi:10.1016/j.jqsrt.2017.01.036}}.

\bibitem{17LyPe.C2H2}
O.~M. Lyulin, V.~I. Perevalov, {ASD-1000: high-resolution, high-temperature
  acetylene spectroscopic databank}, J. Quant. Spectrosc. Radiat. Transf. 201
  (2017) 94--103.
\newblock \href {http://dx.doi.org/10.1016/j.jqsrt.2017.06.032}
  {\path{doi:10.1016/j.jqsrt.2017.06.032}}.

\bibitem{16LyPexx.C2H2}
O.~M. Lyulin, V.~I. Perevalov, {Global modeling of vibration-rotation spectra
  of the acetylene molecule}, J. Quant. Spectrosc. Radiat. Transf. 177 (2016)
  59--74.
\newblock \href {http://dx.doi.org/10.1016/j.jqsrt.2015.12.021}
  {\path{doi:10.1016/j.jqsrt.2015.12.021}}.

\bibitem{jt412}
T.~Furtenbacher, A.~G. {Cs\'asz\'ar}, J.~Tennyson, {MARVEL: measured active
  rotational-vibrational energy levels}, J. Mol. Spectrosc. 245 (2007)
  115--125.

\bibitem{12FuCsxx.marvel}
T.~Furtenbacher, A.~G. {Cs\'asz\'ar}, The role of intensities in determining
  characteristics of spectroscopic networks, J. Molec. Struct. (THEOCHEM) 1009
  (2012) 123 -- 129.
\newblock \href
  {http://dx.doi.org/http://dx.doi.org/10.1016/j.molstruc.2011.10.057}
  {\path{doi:http://dx.doi.org/10.1016/j.molstruc.2011.10.057}}.

\bibitem{11CsFuxx.marvel}
A.~G. {Cs\'asz\'ar}, T.~Furtenbacher, Spectroscopic networks, J. Mol.
  Spectrosc. 266 (2011) 99 -- 103.
\newblock \href {http://dx.doi.org/http://dx.doi.org/10.1016/j.jms.2011.03.031}
  {\path{doi:http://dx.doi.org/10.1016/j.jms.2011.03.031}}.

\bibitem{16ArPeFu.marvel}
P.~{\'A}rend{\'a}s, T.~Furtenbacher, A.~G. Cs{\'a}sz{\'a}r, On spectra of
  spectra, J. Math. Chem. 54 (2016) 806--822.

\bibitem{76FlCaMa}
J.-M. Flaud, C.~Camy-Peyret, J.-P. Maillard, Higher ro-vibrational levels of
  {H2O} deduced from high-resolution oxygen-hydrogen flame spectra between
  2800-6200 cm-1, Mol. Phys. {32} ({1976}) 499--521.
\newblock \href {http://dx.doi.org/10.1080/00268977600103251}
  {\path{doi:10.1080/00268977600103251}}.

\bibitem{04Watson}
J.~K. Watson, Robust weighting in least-squares fits, J. Mol. Spectrosc.
  219~(2) (2003) 326 -- 328.
\newblock \href
  {http://dx.doi.org/http://dx.doi.org/10.1016/S0022-2852(03)00100-0}
  {\path{doi:http://dx.doi.org/10.1016/S0022-2852(03)00100-0}}.

\bibitem{94Watson}
J.~K.~G. Watson, The use of term-value fits in testing spectroscopic
  assignments, J. Mol. Spectrosc. {165} ({1994}) 283--290.
\newblock \href {http://dx.doi.org/10.1006/jmsp.1994.1130}
  {\path{doi:10.1006/jmsp.1994.1130}}.

\bibitem{jt672}
L.~K. McKemmish, T.~Masseron, S.~Sheppard, E.~Sandeman, Z.~Schofield,
  T.~Furtenbacher, A.~G. {Cs\'asz\'ar}, J.~Tennyson, C.~Sousa-Silva, {MARVEL
  analysis of the measured high-resolution spectra of $^{48}$Ti$^{16}$O},
  Astrophys. J. Suppl. 228 (2017) 15.
\newblock \href {http://dx.doi.org/10.3847/1538-4365/228/2/15}
  {\path{doi:10.3847/1538-4365/228/2/15}}.

\bibitem{jt608}
A.~R. {Al Derzi}, T.~Furtenbacher, S.~N. Yurchenko, J.~Tennyson, A.~G.
  Cs\'asz\'ar, {MARVEL analysis of the measured high-resolution spectra of
  $^{14}$NH$_3$}, J. Quant. Spectrosc. Radiat. Transf. 161 (2015) 117--130.
\newblock \href {http://dx.doi.org/10.1016/j.jqsrt.2015.03.034}
  {\path{doi:10.1016/j.jqsrt.2015.03.034}}.

\bibitem{jtNH3update}
T.~Furtenbacher, P.~A. Coles, J.~Tennyson, A.~G. Cs\'asz\'ar, {Updated MARVEL
  energy levels for ammonia}, J. Quant. Spectrosc. Radiat. Transf.

\bibitem{jt454}
J.~Tennyson, P.~F. Bernath, L.~R. Brown, A.~Campargue, M.~R. Carleer, A.~G.
  Cs\'asz\'ar, R.~R. Gamache, J.~T. Hodges, A.~Jenouvrier, O.~V. Naumenko,
  O.~L. Polyansky, L.~S. Rothman, R.~A. Toth, A.~C. Vandaele, N.~F. Zobov,
  L.~Daumont, A.~Z. Fazliev, T.~Furtenbacher, I.~E. Gordon, S.~N. Mikhailenko,
  S.~V. Shirin, {IUPAC critical Evaluation of the Rotational-Vibrational
  Spectra of Water Vapor. Part I. Energy Levels and Transition Wavenumbers for
  H$_2$$^{17}$O and H$_2$$^{18}$O}, J. Quant. Spectrosc. Radiat. Transf. 110
  (2009) 573--596.
\newblock \href {http://dx.doi.org/10.1016/j.jqsrt.2009.02.014}
  {\path{doi:10.1016/j.jqsrt.2009.02.014}}.

\bibitem{jt482}
J.~Tennyson, P.~F. Bernath, L.~R. Brown, A.~Campargue, M.~R. Carleer, A.~G.
  Cs\'asz\'ar, L.~Daumont, R.~R. Gamache, J.~T. Hodges, O.~V. Naumenko, O.~L.
  Polyansky, L.~S. Rothman, R.~A. Toth, A.~C. Vandaele, N.~F. Zobov, A.~Z.
  Fazliev, T.~Furtenbacher, I.~E. Gordon, S.~N. Mikhailenko, B.~A. Voronin,
  {IUPAC critical Evaluation of the Rotational-Vibrational Spectra of Water
  Vapor. Part II. Energy Levels and Transition Wavenumbers for HD$^{16}$O,
  HD$^{17}$O, and HD$^{18}$O}, J. Quant. Spectrosc. Radiat. Transf. 111 (2010)
  2160--2184.
\newblock \href {http://dx.doi.org/10.1016/j.jqsrt.2010.06.012}
  {\path{doi:10.1016/j.jqsrt.2010.06.012}}.

\bibitem{jt539}
J.~Tennyson, P.~F. Bernath, L.~R. Brown, A.~Campargue, M.~R. Carleer, A.~G.
  Cs\'asz\'ar, L.~Daumont, R.~R. Gamache, J.~T. Hodges, O.~V. Naumenko, O.~L.
  Polyansky, L.~S. Rothmam, A.~C. Vandaele, N.~F. Zobov, A.~R. {Al Derzi},
  C.~F\'abri, A.~Z. Fazliev, T.~Furtenbacher, I.~E. Gordon, L.~Lodi, I.~I.
  Mizus, {IUPAC critical evaluation of the rotational-vibrational spectra of
  water vapor. Part III. Energy levels and transition wavenumbers for
  H$_2$$^{16}$O}, J. Quant. Spectrosc. Radiat. Transf. 117 (2013) 29--80.
\newblock \href {http://dx.doi.org/10.1016/j.jqsrt.2012.10.002}
  {\path{doi:10.1016/j.jqsrt.2012.10.002}}.

\bibitem{jt576}
J.~Tennyson, P.~F. Bernath, L.~R. Brown, A.~Campargue, A.~G. Cs\'asz\'ar,
  L.~Daumont, R.~R. Gamache, J.~T. Hodges, O.~V. Naumenko, O.~L. Polyansky,
  L.~S. Rothmam, A.~C. Vandaele, N.~F. Zobov, N.~D\'enes, A.~Z. Fazliev,
  T.~Furtenbacher, I.~E. Gordon, S.-M. Hu, T.~Szidarovszky, I.~A. Vasilenko,
  {IUPAC critical evaluation of the rotational-vibrational spectra of water
  vapor. Part IV. Energy levels and transition wavenumbers for D$_2$$^{16}$O,
  D$_2$$^{17}$O and D$_2$$^{18}$O}, J. Quant. Spectrosc. Radiat. Transf. 142
  (2014) 93--108.
\newblock \href {http://dx.doi.org/10.1016/j.jqsrt.2014.03.019}
  {\path{doi:10.1016/j.jqsrt.2014.03.019}}.

\bibitem{jtwaterupdate}
T.~Furtenbacher, N.~D{\'e}nes, J.~Tennyson, O.~V. Naumenko, O.~L. Polyansky,
  N.~F. Zobov, A.~G. Cs\'asz\'ar, {The 2016 Update of the IUPAC Database of
  Water Energy Levels}, J. Quant. Spectrosc. Radiat. Transf.(in preparation).

\bibitem{13FuSzFa.marvel}
T.~Furtenbacher, T.~Szidarovszky, C.~F{\'a}bri, A.~G. Cs{\'a}sz{\'a}r, {MARVEL
  analysis of the rotational--vibrational states of the molecular ions
  H$_2$D$^+$ and D$_2$H$^+$}, Phys. Chem. Chem. Phys. 15 (2013) 10181--10193.

\bibitem{13FuSzMa.marvel}
T.~Furtenbacher, T.~Szidarovszky, E.~M{\'a}tyus, C.~F{\'a}bri, A.~G.
  Cs{\'a}sz{\'a}r, {Analysis of the Rotational--Vibrational States of the
  Molecular Ion H$_3^+$}, J. Chem. Theory Comput. 9 (2013) 5471--5478.

\bibitem{jt637}
T.~Furtenbacher, I.~Szab{\'o}, A.~G. Cs{\'a}sz{\'a}r, P.~F. Bernath, S.~N.
  Yurchenko, J.~Tennyson, Experimental energy levels and partition function of
  the 12c2 molecule, Astrophys. J. Suppl. 224 (2016) 44.
\newblock \href {http://dx.doi.org/10.3847/0067-0049/224/2/44}
  {\path{doi:10.3847/0067-0049/224/2/44}}.

\bibitem{75BrHoHu.linear}
J.~M. Brown, J.~T. Hougen, K.-P. Huber, J.~W.~C. Johns, I.~Kopp,
  H.~Lefebvre-Brion, A.~J. Merer, D.~A. Ramsay, J.~Rostas, R.~N. Zare, {The
  labeling of parity doublet levels in linear molecules}, J. Mol. Spectrosc. 55
  (1975) 500--503.
\newblock \href {http://dx.doi.org/10.1016/0022-2852(75)90291-X}
  {\path{doi:10.1016/0022-2852(75)90291-X}}.

\bibitem{72Plxxxxa.C2H2}
J.~Pl\'iva, {Spectrum of acetylene in the 5-micron region}, J. Mol. Spectrosc.
  44 (1972) 145 -- 164.
\newblock \href {http://dx.doi.org/10.1016/0022-2852(72)90198-1}
  {\path{doi:10.1016/0022-2852(72)90198-1}}.

\bibitem{72WiWi.HCNO}
M.~Winnewisser, B.~P. Winnewisser,
  \href{http://www.sciencedirect.com/science/article/pii/0022285272901294}{{Mi%
llimeter wave rotational spectrum of HCNO in vibrationally excited states}}, J.
  Mol. Spectrosc. 41~(1) (1972) 143 -- 176.
\newblock \href
  {http://dx.doi.org/https://doi.org/10.1016/0022-2852(72)90129-4}
  {\path{doi:https://doi.org/10.1016/0022-2852(72)90129-4}}.
\newline\urlprefix\url{http://www.sciencedirect.com/science/article/pii/002228%
5272901294}

\bibitem{99HeLiVa.linear}
M.~Herman, J.~Lievin, J.~Vander~Auwera, A.~Campargue, {Global and Accurate
  Vibration Hamiltonians from High-Resolution Molecular Spectroscopy}, Vol. 108
  of Adv. Chem. Phys., Wiley and Sons, Inc., New York, NY, 1999.

\bibitem{82HeLixx.C2H2}
M.~Herman, J.~Lievin, {Acetylene- From intensity alternation in spectra to
  ortho and para molecule}, J. Chem. Educ. 59 (1982) 17.
\newblock \href {http://dx.doi.org/10.1021/ed059p17}
  {\path{doi:10.1021/ed059p17}}.

\bibitem{98BuJexx.method}
P.~R. Bunker, P.~Jensen, Molecular Symmetry and Spectroscopy, 2nd Edition, NRC
  Research Press, Ottawa, 1998.

\bibitem{jt562}
J.~Tennyson, P.~F. Bernath, L.~R. Brown, A.~Campargue, A.~G. Cs\'asz\'ar,
  L.~Daumont, R.~R. Gamache, J.~T. Hodges, O.~V. Naumenko, O.~L. Polyansky,
  L.~S. Rothman, A.~C. Vandaele, N.~F. Zobov, {A Database of Water Transitions
  from Experiment and Theory (IUPAC Technical Report)}, Pure Appl. Chem. 86
  (2014) 71--83.
\newblock \href {http://dx.doi.org/10.1515/pac-2014-5012}
  {\path{doi:10.1515/pac-2014-5012}}.

\bibitem{09YuDrPe.C2H2}
S.~Yu, B.~J. Drouin, J.~C. Pearson, {Terahertz Spectroscopy of the Bending
  Vibrations of Acetylene $^{12}$C$_2$H$_2$}, Astrophys. J. 705 (2009)
  786--790.
\newblock \href {http://dx.doi.org/10.1088/0004-637X/705/1/786}
  {\path{doi:10.1088/0004-637X/705/1/786}}.

\bibitem{91KaHeDi.C2H2}
Y.~Kabbadj, M.~Herman, G.~Dilonardo, L.~Fusina, J.~W.~C. Johns, {The bending
  energy-levels of C$_2$H$_2$}, J. Mol. Spectrosc. 150 (1991) 535--565.
\newblock \href {http://dx.doi.org/10.1016/0022-2852(91)90248-9}
  {\path{doi:10.1016/0022-2852(91)90248-9}}.

\bibitem{10AmHeFa.C2H2}
B.~Amyay, M.~Herman, A.~Fayt, L.~Fusina, A.~Predoi-Cross, {High resolution FTIR
  investigation of $^{12}$C$_2$H$_2$ in the FIR spectral range using
  synchrotron radiation}, Chem. Phys. Lett. 491 (2010) 17--19.
\newblock \href {http://dx.doi.org/10.1016/j.cplett.2010.03.053}
  {\path{doi:10.1016/j.cplett.2010.03.053}}.

\bibitem{11DrYuxx.C2H2}
B.~J. Drouin, S.~Yu, {Acetylene spectra near 2.6 THz}, J. Mol. Spectrosc. 269
  (2011) 254--256.
\newblock \href {http://dx.doi.org/10.1016/j.jms.2011.06.004}
  {\path{doi:10.1016/j.jms.2011.06.004}}.

\bibitem{17JaLyPe.C2H2}
D.~Jacquemart, O.~M. Lyulin, V.~I. Perevalov, {Recommended acetylene line list
  in the 20 -- 240 cm$^{-1}$ and 400 -- 630 cm$^{-1}$ regions: new measurements
  and global modeling}, J. Quant. Spectrosc. Radiat. Transf.\href
  {http://dx.doi.org/10.1016/j.jqsrt.2017.03.008}
  {\path{doi:10.1016/j.jqsrt.2017.03.008}}.

\bibitem{81HiKaxx.C2H2}
J.~Hietanen, J.~Kauppinen, {High-resolution infrared-spectrum of acetylene in
  the region of the bending fundamental $\nu_5$}, Mol. Phys. 42 (1981)
  411--423.
\newblock \href {http://dx.doi.org/10.1080/00268978100100351}
  {\path{doi:10.1080/00268978100100351}}.

\bibitem{93WeBlNa.C2H2}
M.~Weber, W.~E. Blass, S.~Nadler, G.~W. Halsey, W.~C. Maguire, J.~J. Hillman,
  {Resonance effects in C$_2$H$_2$ near 13.7 $\mu$m. Part H: The two quantum
  hotbands}, Spectra Chimica Acta A 49 (1993) 1659 -- 1681.
\newblock \href {http://dx.doi.org/10.1016/0584-8539(93)80124-S}
  {\path{doi:10.1016/0584-8539(93)80124-S}}.

\bibitem{00MaDaCl.C2H2}
J.~Y. Mandin, V.~Dana, C.~Claveau, {Line intensities in the $\nu_5$ band of
  acetylene $^{12}$C$_2$H$_2$}, J. Quant. Spectrosc. Radiat. Transf. 67 (2000)
  429--446.
\newblock \href {http://dx.doi.org/10.1016/S0022-4073(00)00010-8}
  {\path{doi:10.1016/S0022-4073(00)00010-8}}.

\bibitem{01JaClMa.C2H2}
D.~Jacquemart, C.~Claveau, J.~Y. Mandin, V.~Dana, {Line intensities of hot
  bands in the 13.6 $\mu$m spectral region of acetylene $^{12}$C$_2$H$_2$}, J.
  Quant. Spectrosc. Radiat. Transf. 69 (2001) 81--101.
\newblock \href {http://dx.doi.org/10.1016/S0022-4073(00)00063-7}
  {\path{doi:10.1016/S0022-4073(00)00063-7}}.

\bibitem{50BeNixx.C2H2}
E.~E. Bell, H.~H. Nielsen, {The infra-red spectrum of Acetylene}, J. Chem.
  Phys. 18 (1950) 1382--1394.
\newblock \href {http://dx.doi.org/10.1063/1.1747483}
  {\path{doi:10.1063/1.1747483}}.

\bibitem{10GoJaLa.C2H2}
L.~Gomez, D.~Jacquemart, N.~Lacome, J.-Y. Mandin, {New line intensity
  measurements for $^{12}$C$_2$H$_2$ around 7.7 $\mu$m and HITRAN format line
  list for applications}, J. Quant. Spectrosc. Radiat. Transf. 111 (2010)
  2256--2264.
\newblock \href {http://dx.doi.org/10.1016/j.jqsrt.2010.01.031}
  {\path{doi:10.1016/j.jqsrt.2010.01.031}}.

\bibitem{09GoJaLa.C2H2}
L.~Gomez, D.~Jacquemart, N.~Lacome, J.~Y. Mandin, {Line intensities of
  $^{12}$C$_2$H$_2$ in the 7.7 $\mu$m spectral region}, J. Quant. Spectrosc.
  Radiat. Transf. 110 (2009) 2102--2114.
\newblock \href {http://dx.doi.org/10.1016/j.jqsrt.2009.05.018}
  {\path{doi:10.1016/j.jqsrt.2009.05.018}}.

\bibitem{00Vander.C2H2}
J.~Vander~Auwera, {Absolute intensities measurements in the $\nu_4+\nu_5$ band
  of $^{12}$C$_2$H$_2$: Analysis of Herman-Wallis effects and forbidden
  transitions}, J. Mol. Spectrosc. 201 (2000) 143--150.
\newblock \href {http://dx.doi.org/10.1006/jmsp.2000.8079}
  {\path{doi:10.1006/jmsp.2000.8079}}.

\bibitem{09AmRoHe.C2H2}
B.~Amyay, S.~Robert, M.~Herman, A.~Fayt, B.~Raghavendra, A.~Moudens,
  J.~Thievin, B.~Rowe, R.~Georges, {Vibration-rotation pattern in acetylene.
  II. Introduction of Coriolis coupling in the global model and analysis of
  emission spectra of hot acetylene around 3 $\mu$m}, J. Chem. Phys. 131 (2009)
  114301.
\newblock \href {http://dx.doi.org/10.1063/1.3200928}
  {\path{doi:10.1063/1.3200928}}.

\bibitem{03JaMaDab.C2H2}
D.~Jacquemart, J.~Y. Mandin, V.~Dana, L.~Regalia-Jarlot, J.~Plateaux,
  D.~Decatoire, L.~S. Rothman, {The spectrum of acetylene in the 5-$\mu$m
  region from new line-parameter measurements}, J. Quant. Spectrosc. Radiat.
  Transf. 76 (2003) 237--267.
\newblock \href {http://dx.doi.org/10.1016/S0022-4073(02)00055-9}
  {\path{doi:10.1016/S0022-4073(02)00055-9}}.

\bibitem{03JaMaDa.C2H2}
D.~Jacquemart, J.~Y. Mandin, V.~Dana, C.~Claveau, J.~Vander~Auwera, A.~Herman,
  L.~S. Rothman, L.~Regalia-Jarlot, A.~Barbe, {The IR acetylene spectrum in
  HITRAN: update and new results}, J. Quant. Spectrosc. Radiat. Transf. 82
  (2003) 363--382.
\newblock \href {http://dx.doi.org/10.1016/S0022-4073(03)00163-8}
  {\path{doi:10.1016/S0022-4073(03)00163-8}}.

\bibitem{02JaMaDa.C2H2}
D.~Jacquemart, J.~Y. Mandin, V.~Dana, L.~Regalia-Jarlot, X.~Thomas, P.~Von~der
  Heyden, {Multispectrum fitting measurements of line parameters for 5-$\mu$m
  cold bands of acetylene}, J. Quant. Spectrosc. Radiat. Transf. 75 (2002)
  397--422.
\newblock \href {http://dx.doi.org/10.1016/S0022-4073(02)00017-1}
  {\path{doi:10.1016/S0022-4073(02)00017-1}}.

\bibitem{98BeCaLo.C2H2}
D.~Bermejo, P.~Cancio, G.~Di~Lonardo, L.~Fusina, {High resolution Raman
  spectroscopy from vibrationally excited states populated by a stimulated
  Raman process: $2\nu_2-\nu_2$ of $^{12}$C$_2$H$_2$ and $^{12}$C$_2$H$_2$}, J.
  Chem. Phys. 108 (1998) 7224--7228.
\newblock \href {http://dx.doi.org/10.1063/1.476140}
  {\path{doi:10.1063/1.476140}}.

\bibitem{07JaLaMa.C2H2}
D.~Jacquemart, N.~Lacome, J.~Y. Mandin, V.~Dana, O.~M. Lyulin, V.~I. Perevalov,
  {Multispectrum fitting of line parameters for $^{12}$C$_2$H$_2$ in the
  3.8-$\mu$m spectral region}, J. Quant. Spectrosc. Radiat. Transf. 103 (2007)
  478--495.
\newblock \href {http://dx.doi.org/10.1016/j.jqsrt.2006.06.008}
  {\path{doi:10.1016/j.jqsrt.2006.06.008}}.

\bibitem{72PaMiNa.C2H2}
K.~F. Palmer, M.~E. Mickelson, K.~Narahari~Rao, {Investigations of several
  infrared bands of $^{12}$C$_{2}$H $_{2}$ and studies of the effects of
  vibrational rotational interactions}, J. Mol. Spectrosc. 44 (1972) 131--144.
\newblock \href {http://dx.doi.org/10.1016/0022-2852(72)90197-X}
  {\path{doi:10.1016/0022-2852(72)90197-X}}.

\bibitem{93VaHuCa.C2H2}
J.~{Vander Auwera}, D.~Hurtmans, M.~Carleer, M.~Herman, {The $\nu_3$
  Fundamental in C$_2$H$_2$}, J. Mol. Spectrosc. 157 (1993) 337 -- 357.
\newblock \href {http://dx.doi.org/10.1006/jmsp.1993.1027}
  {\path{doi:10.1006/jmsp.1993.1027}}.

\bibitem{93DcSaJo.C2H2}
R.~Dcunha, Y.~A. Sarma, V.~A. Job, G.~Guelachvili, K.~N. Rao, {Fermi Coupling
  and $\ell$-Type Resonance Effects in the Hot Bands of Acetylene: The
  2650-cm$^{-1}$ Region}, J. Mol. Spectrosc. 157 (1993) 358 -- 368.
\newblock \href {http://dx.doi.org/10.1006/jmsp.1993.1028}
  {\path{doi:10.1006/jmsp.1993.1028}}.

\bibitem{95SaDcGub.C2H2}
Y.~A. Sarma, R.~Dcunha, G.~Guelachvili, R.~Farrenq, V.~M. Devi, D.~C. Benner,
  K.~N. Rao, {Stretch-bend levels of acetylene - analysis of the hot bands in
  the 3300cm$^{-1}$ region}, J. Mol. Spectrosc. 173 (1995) 574--584.
\newblock \href {http://dx.doi.org/10.1006/jmsp.1995.1258}
  {\path{doi:10.1006/jmsp.1995.1258}}.

\bibitem{06LyPeMa.C2H2}
O.~M. Lyulin, V.~I. Perevalov, J.~Y. Mandin, V.~Dana, D.~Jacquemart,
  L.~Regalia-Jarlot, A.~Barbe, {Line intensities of acetylene in the 3-$\mu$m
  region: New measurements of weak hot bands and global fitting}, J. Quant.
  Spectrosc. Radiat. Transf. 97 (2006) 81--98.
\newblock \href {http://dx.doi.org/10.1016/j.jqsrt.2004.12.022}
  {\path{doi:10.1016/j.jqsrt.2004.12.022}}.

\bibitem{05MaJaDa.C2H2}
J.~Y. Mandin, D.~Jacquemart, V.~Dana, L.~Regalia-Jarlot, A.~Barbe, {Line
  intensities of acetylene at 3 $\mu$m}, J. Quant. Spectrosc. Radiat. Transf.
  92 (2005) 239--260.
\newblock \href {http://dx.doi.org/10.1016/j.jqsrt.2004.07.024}
  {\path{doi:10.1016/j.jqsrt.2004.07.024}}.

\bibitem{95SaDcGu.C2H2}
Y.~A. Sarma, R.~Dcunha, G.~Guelachvili, R.~Farrenq, K.~N. Rao, {Stretch-bend
  levels of acetylene - analysis of the hot bands in the 3800cm$^{-1}$ region},
  J. Mol. Spectrosc. 173 (1995) 561--573.
\newblock \href {http://dx.doi.org/10.1006/jmsp.1995.1257}
  {\path{doi:10.1006/jmsp.1995.1257}}.

\bibitem{99BeMaLo.C2H2}
D.~Bermejo, R.~Z. Martinez, G.~Di~Lonardo, L.~Fusina, {High resolution Raman
  spectroscopy from vibrationally excited states populated by a stimulated
  Raman process. Transitions from $v_2=1$ in $^{12}$C$_2$H$_2$ and
  $^{13}$C$_2$H$_2$}, J. Chem. Phys. 111 (1999) 519--524.
\newblock \href {http://dx.doi.org/10.1063/1.479331}
  {\path{doi:10.1063/1.479331}}.

\bibitem{07LyPeMa.C2H2}
O.~M. Lyulin, V.~I. Perevalov, J.~Y. Mandin, V.~Dana, F.~Gueye, X.~Thomas,
  P.~Von~der Heyden, D.~Decatoire, L.~Regalia-Jarlot, D.~Jacquemart, N.~Lacome,
  {Line intensities of acetylene: Measurements in the 2.5-$\mu$m spectral
  region and global modeling in the $\Delta p$ = 4 and 6 series}, J. Quant.
  Spectrosc. Radiat. Transf. 103 (2007) 496--523.
\newblock \href {http://dx.doi.org/10.1016/j.jqsrt.2006.07.002}
  {\path{doi:10.1016/j.jqsrt.2006.07.002}}.

\bibitem{06GiFaSo.C2H2}
V.~Girard, R.~Farrenq, E.~Sorokin, I.~T. Sorokina, G.~Guelachvili, N.~Picque,
  {Acetylene weak bands at 2.5 $\mu$m from intracavity Cr$^{2+}$: ZnSe laser
  absorption observed with time-resolved Fourier transform spectroscopy}, Chem.
  Phys. Lett. 419 (2006) 584--588.
\newblock \href {http://dx.doi.org/10.1016/j.cplett.2005.12.029}
  {\path{doi:10.1016/j.cplett.2005.12.029}}.

\bibitem{91DcSaGu.C2H2}
R.~Dcuhna, Y.~A. Sarma, G.~Guelachvili, R.~Farrenq, Q.~L. Kou, V.~M. Devi,
  D.~C. Benner, K.~N. Rao, {Analysis of the high-resolution spectrum of
  acetylene in the {2.4 $\mu$m} region}, J. Mol. Spectrosc. 148 (1991)
  213--225.
\newblock \href {http://dx.doi.org/10.1016/0022-2852(91)90048-F}
  {\path{doi:10.1016/0022-2852(91)90048-F}}.

\bibitem{72BaGhNa.C2H2}
A.~Baldacci, S.~Ghersetti, K.~N. Rao, {Assignments of the {$^{12}$C$_2$H$_2$}
  bands at 2.1-2.2 $\mu$m}, J. Mol. Spectrosc. 41 (1972) 222 -- 225.
\newblock \href {http://dx.doi.org/10.1016/0022-2852(72)90134-8}
  {\path{doi:10.1016/0022-2852(72)90134-8}}.

\bibitem{07LyPeGu.C2H2}
O.~M. Lyulin, V.~I. Perevalov, F.~Gueye, J.~Y. Mandin, V.~Dana, X.~Thomas,
  P.~Von~der Heyden, L.~Regalia-Jarlot, A.~Barbe, {Line positions and
  intensities of acetylene in the 2.2-$\mu$m region}, J. Quant. Spectrosc.
  Radiat. Transf. 104 (2007) 133--154.
\newblock \href {http://dx.doi.org/10.1016/j.jqsrt.2006.08.018}
  {\path{doi:10.1016/j.jqsrt.2006.08.018}}.

\bibitem{08LyJaLa.C2H2}
O.~M. Lyulin, D.~Jacquemart, N.~Lacome, V.~I. Perevalov, J.~Y. Mandin, {Line
  parameters of acetylene in the 1.9 and 1.7 $\mu$m spectral regions}, J.
  Quant. Spectrosc. Radiat. Transf. 109 (2008) 1856--1874.
\newblock \href {http://dx.doi.org/10.1016/j.jqsrt.2007.11.016}
  {\path{doi:10.1016/j.jqsrt.2007.11.016}}.

\bibitem{96KeMeKl.C2H2}
K.~A. Keppler, G.~C. Mellau, S.~Klee, B.~P. Winnewisser, M.~Winnewisser,
  J.~Pliva, K.~N. Rao, {Precision measurements of acetylene spectra at 1.4-1.7
  $\mu$m recorded with 352.5-m pathlength}, J. Mol. Spectrosc. 175 (1996)
  411--420.
\newblock \href {http://dx.doi.org/10.1006/jmsp.1996.0047}
  {\path{doi:10.1006/jmsp.1996.0047}}.

\bibitem{08RoHeFa.C2H2}
S.~Robert, M.~Herman, A.~Fayt, A.~Campargue, S.~Kassi, A.~Liu, L.~Wang,
  G.~Di~Lonardo, L.~Fusina, {Acetylene, $^{12}$C$_2$H$_2$: new CRDS data and
  global vibration-rotation analysis up to 8600 cm$^{-1}$}, Mol. Phys. 106
  (2008) 2581--2605.
\newblock \href {http://dx.doi.org/10.1080/00268970802620709}
  {\path{doi:10.1080/00268970802620709}}.

\bibitem{07TrMaDa.C2H2}
H.~Tran, J.~Y. Mandin, V.~Dana, L.~Regalia-Jarlot, X.~Thomas, P.~Von~der
  Heyden, {Line intensities in the 1.5-$\mu$m spectral region of acetylene}, J.
  Quant. Spectrosc. Radiat. Transf. 108 (2007) 342--362.
\newblock \href {http://dx.doi.org/10.1016/j.jqsrt.2007.04.008}
  {\path{doi:10.1016/j.jqsrt.2007.04.008}}.

\bibitem{09LyPeTr.C2H2}
O.~M. Lyulin, V.~I. Perevalov, H.~Tran, J.~Y. Mandin, V.~Dana,
  L.~Regalia-Jarlot, X.~Thomas, D.~Decatoire, {Line intensities of acetylene:
  New measurements in the 1.5-$\mu$m spectral region and global modelling in
  the $\Delta P=10$ series}, J. Quant. Spectrosc. Radiat. Transf. 110 (2009)
  1815--1824.
\newblock \href {http://dx.doi.org/10.1016/j.jqsrt.2009.04.012}
  {\path{doi:10.1016/j.jqsrt.2009.04.012}}.

\bibitem{16KaNaVa.C2H2}
J.~Karhu, J.~Nauta, M.~Vainio, M.~Mets{\"a}l{\"a}, S.~Hoekstra, L.~Halonen,
  {Double resonant absorption measurement of acetylene symmetric vibrational
  states probed with cavity ring down spectroscopy}, J. Chem. Phys. 144 (2016)
  244201.
\newblock \href {http://dx.doi.org/10.1063/1.4954159}
  {\path{doi:10.1063/1.4954159}}.

\bibitem{94KoGuTe.C2H2}
Q.~Kou, G.~Guelachvili, M.~A. Temsamani, M.~Herman, {The absorption spectrum of
  {C$_2$H$_2$} around $\nu_1+\nu_3$ - energy standards in the 1.5 $\mu$m region
  and vibrational clustering}, Can. J. Phys. 72 (1994) 1241--1250.
\newblock \href {http://dx.doi.org/10.1139/p94-160}
  {\path{doi:10.1139/p94-160}}.

\bibitem{15TwCiSe.C2H2}
S.~Twagirayezu, M.~J. Cich, T.~J. Sears, C.~P. McRaven, G.~E. Hall,
  {Frequency-comb referenced spectroscopy of $v_4$- and $v_5$-excited hot bands
  in the 1.5 $\mu$m spectrum of C$_2$H$_2$}, J. Mol. Spectrosc. 316 (2015)
  64--71.
\newblock \href {http://dx.doi.org/10.1016/j.jms.2015.06.010}
  {\path{doi:10.1016/j.jms.2015.06.010}}.

\bibitem{02HaAuxx.C2H2}
R.~El~Hachtouki, J.~Vander~Auwera, {Absolute line intensities in acetylene: The
  1.5-$\mu$m region}, J. Mol. Spectrosc. 216 (2002) 355--362.
\newblock \href {http://dx.doi.org/10.1006/jmsp.2002.8660}
  {\path{doi:10.1006/jmsp.2002.8660}}.

\bibitem{77BaGhNa.C2H2}
A.~Baldacci, S.~Ghersetti, K.~N. Rao, {Interpretation of the acetylene spectrum
  at 1.5 $\mu$m}, J. Mol. Spectrosc. 68 (1977) 183--194.
\newblock \href {http://dx.doi.org/10.1016/0022-2852(77)90436-2}
  {\path{doi:10.1016/0022-2852(77)90436-2}}.

\bibitem{05EdBaMa.C2H2}
C.~S. Edwards, G.~P. Barwood, H.~S. Margolis, P.~Gill, W.~R.~C. Rowley,
  {High-precision frequency measurements of the $v_1 + v_3$ combination band of
  $^{12}$C$_2$H$_2$ in the 1.5 $\mu$m region}, J. Mol. Spectrosc. 234 (2005)
  143--148.
\newblock \href {http://dx.doi.org/10.1016/j.jms.2005.08.014}
  {\path{doi:10.1016/j.jms.2005.08.014}}.

\bibitem{13ZoGiBa.C2H2}
A.~M. Zolot, F.~R. Giorgetta, E.~Baumann, W.~C. Swann, I.~Coddington, N.~R.
  Newbury, {Broad-band frequency references in the near-infrared: Accurate dual
  comb spectroscopy of methane and acetylene}, J. Quant. Spectrosc. Radiat.
  Transf. 118 (2013) 26--39.
\newblock \href {http://dx.doi.org/10.1016/j.jqsrt.2012.11.024}
  {\path{doi:10.1016/j.jqsrt.2012.11.024}}.

\bibitem{00MoDuJa.C2H2}
D.~B. Moss, Z.~C. Duan, M.~P. Jacobson, J.~P. O'Brien, R.~W. Field,
  {Observation of coriolis coupling between $\nu_2+4\nu_4$ and $7\nu_4$ in
  acetylene $^1\Sigma^+_g$ by stimulated emission pumping spectroscopy}, J.
  Mol. Spectrosc. 199 (2000) 265--274.
\newblock \href {http://dx.doi.org/10.1006/jmsp.1999.7994}
  {\path{doi:10.1006/jmsp.1999.7994}}.

\bibitem{96NaLaAw.C2H2}
K.~Nakagawa, M.~deLabachelerie, Y.~Awaji, M.~Kourogi, {Accurate optical
  frequency atlas of the 1.5-$\mu$m bands of acetylene}, J. Opt. Soc. Am. B 13
  (1996) 2708--2714.
\newblock \href {http://dx.doi.org/10.1364/JOSAB.13.002708}
  {\path{doi:10.1364/JOSAB.13.002708}}.

\bibitem{11AmHeFa.C2H2}
B.~Amyay, M.~Herman, A.~Fayt, A.~Campargue, S.~Kassi, {Acetylene,
  ($^{12}$C$_2$H$_2$): Refined analysis of CRDS spectra around 1.52 $\mu$m}, J.
  Mol. Spectrosc. 267 (2011) 80--91.
\newblock \href {http://dx.doi.org/10.1016/j.jms.2011.02.015}
  {\path{doi:10.1016/j.jms.2011.02.015}}.

\bibitem{15LyVaCa.C2H2}
O.~M. Lyulin, J.~Vander~Auwera, A.~Campargue, {The Fourier transform absorption
  spectrum of acetylene between 7000 and 7500 cm$^{-1}$}, J. Quant. Spectrosc.
  Radiat. Transf. 160 (2015) 85--93.
\newblock \href {http://dx.doi.org/10.1016/j.jqsrt.2015.03.018}
  {\path{doi:10.1016/j.jqsrt.2015.03.018}}.

\bibitem{09JaLaMa.C2H2}
D.~Jacquemart, N.~Lacome, J.~Y. Mandin, V.~Dana, H.~Tran, F.~K. Gueye, O.~M.
  Lyulin, V.~I. Perevalov, L.~Regalia-Jarlot, {The IR spectrum of
  $^{12}$C$_2$H$_2$: Line intensity measurements in the 1.4 $\mu$m region and
  update of the databases}, J. Quant. Spectrosc. Radiat. Transf. 110 (2009)
  717--732.
\newblock \href {http://dx.doi.org/10.1016/j.jqsrt.2008.10.002}
  {\path{doi:10.1016/j.jqsrt.2008.10.002}}.

\bibitem{02VaElBr.C2H2}
J.~Vander~Auwera, R.~El~Hachtouki, L.~R. Brown, {Absolute line wavenumbers in
  the near infrared: $^{12}$C$_2$H$_2$ and $^{12}$C$^{16}$O$_2$}, Mol. Phys.
  100 (2002) 3563--3576.
\newblock \href {http://dx.doi.org/10.1080/00268970210162880}
  {\path{doi:10.1080/00268970210162880}}.

\bibitem{16LyVaCa.C2H2}
O.~M. Lyulin, J.~Vander~Auwera, A.~Campargue, {The Fourier transform absorption
  spectrum of acetylene between 8280 and 8700 cm$^{-1}$}, J. Quant. Spectrosc.
  Radiat. Transf. 177 (2016) 234--240.
\newblock \href {http://dx.doi.org/10.1016/j.jqsrt.2015.11.026}
  {\path{doi:10.1016/j.jqsrt.2015.11.026}}.

\bibitem{17BeLyHu.C2H2}
S.~B\'eguier, O.~M. Lyulin, S.-M. Hu, A.~Campargue, {Line intensity
  measurements for acetylene between 8980 and 9420 cm$^{-1}$}, J. Quant.
  Spectrosc. Radiat. Transf. 189 (2017) 417--420.
\newblock \href {http://dx.doi.org/10.1016/j.jqsrt.2016.12.020}
  {\path{doi:10.1016/j.jqsrt.2016.12.020}}.

\bibitem{89HeHuVe.C2H2}
M.~Herman, T.~R. Huet, M.~Vervloet, {Spectroscopy and vibrational couplings in
  the $3\nu_3$ region of acetylene}, Mol. Phys. 66 (1989) 333--353.
\newblock \href {http://dx.doi.org/10.1080/00268978900100161}
  {\path{doi:10.1080/00268978900100161}}.

\bibitem{93SaKaxx.C2H2}
J.~Sakai, M.~Katayama, {Diode Laser Spectroscopy of Acetylene: The $2\nu_1$ +
  $2\nu_3$ + $\nu_4$ - $\nu_5$ and $4\nu_1$ - $\nu_5$ Interacting Band System},
  J. Mol. Spectrosc. 157 (1993) 532 -- 535.
\newblock \href {http://dx.doi.org/10.1006/jmsp.1993.1042}
  {\path{doi:10.1006/jmsp.1993.1042}}.

\bibitem{03HeKeHu.C2H2}
F.~Herregodts, E.~Kerrinckx, T.~R. Huet, J.~Vander~Auwera, {Absolute line
  intensities in the $\nu_1+3\nu_3$ band of $^{12}$C$_2$H$_2$ by laser
  photoacoustic spectroscopy and Fourier transform spectroscopy}, Mol. Phys.
  101 (2003) 3427--3438.
\newblock \href {http://dx.doi.org/10.1080/00268970310001632426}
  {\path{doi:10.1080/00268970310001632426}}.

\bibitem{92SaKaxx.C2H2}
J.~Sakai, M.~Katayama, {Diode laser spectroscopy of acetylene: $3\nu_3+\nu_1$3
  Region at 0.77 $\mu$m}, J. Mol. Spectrosc. 154 (1992) 277 -- 287.
\newblock \href {http://dx.doi.org/10.1016/0022-2852(92)90208-6}
  {\path{doi:10.1016/0022-2852(92)90208-6}}.

\bibitem{94SaSeKa.C2H2}
J.~Sakai, H.~Segawa, M.~Katayama, {Diode Laser Spectroscopy of the $2\nu_1$ +
  $2\nu_2$ + $\nu_3$ Band of Acetylene}, J. Mol. Spectrosc. 164 (1994) 580 --
  582.
\newblock \href {http://dx.doi.org/10.1006/jmsp.1994.1101}
  {\path{doi:10.1006/jmsp.1994.1101}}.

\bibitem{96TeHeSo.C2H2}
M.~A. Temsamani, M.~Herman, S.~A.~B. Solina, J.~P. O'Brien, R.~W. Field,
  {Highly vibrationally excited $^{12}$C$_2$H$_2$ in the X $^1\Sigma^+_g$
  state: Complementarity of absorption and dispersed fluorescence spectra}, J.
  Chem. Phys. 105 (1996) 11357--11359.
\newblock \href {http://dx.doi.org/10.1063/1.472995}
  {\path{doi:10.1063/1.472995}}.

\bibitem{99ElLiCa.C2H2}
M.~I. El~Idrissi, J.~Lievin, A.~Campargue, M.~Herman, {The vibrational energy
  pattern in acetylene (IV): Updated global vibration constants for
  $^{12}$C$_2$H$_2$}, J. Chem. Phys. 110 (1999) 2074--2086.
\newblock \href {http://dx.doi.org/10.1063/1.477817}
  {\path{doi:10.1063/1.477817}}.

\bibitem{72Plxxxxb.C2H2}
J.~Pl\'iva, {Molecular constants for the bending modes of acetylene
  $^{12}$C$_2$H$_2$}, J. Mol. Spectrosc. 44 (1972) 165 -- 182.
\newblock \href {http://dx.doi.org/10.1016/0022-2852(72)90199-3}
  {\path{doi:10.1016/0022-2852(72)90199-3}}.

\bibitem{02MeYaVa.C2H2}
M.~Mets{\"a}l{\"a}, S.~Yang, O.~Vaittinen, L.~Halonen, {Laser-induced dispersed
  vibration-rotation fluorescence of acetylene: Spectra of ortho and para forms
  and partial trapping of vibrational energy}, J. Chem. Phys. 117 (2002)
  8686--8693.
\newblock \href {http://dx.doi.org/10.1063/1.1513464}
  {\path{doi:10.1063/1.1513464}}.

\bibitem{01MeYaVa.C2H2}
M.~Mets{\"a}l{\"a}, S.~F. Yang, A.~Vaittinen, D.~Permogorov, L.~Halonen,
  {High-resolution cavity ring-down study of acetylene between 12260 and 12380
  cm$^{-1}$}, Chem. Phys. Lett. 346 (2001) 373--378.
\newblock \href {http://dx.doi.org/10.1016/S0009-2614(01)00945-9}
  {\path{doi:10.1016/S0009-2614(01)00945-9}}.

\bibitem{99SaPeHa.C2H2}
M.~Saarinen, D.~Permogorov, L.~Halonen, {Collision-induced vibration-rotation
  fluorescence spectra and rovibrational symmetry changes in acetylene}, J.
  Chem. Phys. 110 (1999) 1424--1428.
\newblock \href {http://dx.doi.org/10.1063/1.478017}
  {\path{doi:10.1063/1.478017}}.

\bibitem{97JuHaxx.C2H2}
P.~Jungner, L.~Halonen, {Laser induced vibration-rotation fluorescence and
  infrared forbidden transitions in acetylene}, J. Chem. Phys. 107 (1997)
  1680--1682.
\newblock \href {http://dx.doi.org/10.1063/1.474521}
  {\path{doi:10.1063/1.474521}}.

\bibitem{93ZhHaxx.C2H2}
X.~W. Zhan, L.~Halonen, {High-resolution photoacoustic study of the
  $\nu_1+3\nu_3$ band system of acetylene with a titanium-sapphire ring laser},
  J. Mol. Spectrosc. 160 (1993) 464--470.
\newblock \href {http://dx.doi.org/10.1006/jmsp.1993.1193}
  {\path{doi:10.1006/jmsp.1993.1193}}.

\bibitem{93ZhVaHa.C2H2}
X.~W. Zhan, O.~Vaittinen, L.~Halonen, {High-resolution photoacoustic study of
  acetylene between 11500 and 11900cm$^{-1}$ using a titanium-sapphire ring
  laser}, J. Mol. Spectrosc. 160 (1993) 172--180.
\newblock \href {http://dx.doi.org/10.1006/jmsp.1993.1165}
  {\path{doi:10.1006/jmsp.1993.1165}}.

\bibitem{91ZhVaKa.C2H2}
X.-W. Zhan, O.~Vaittinen, E.~Kauppi, L.~Halonen, {High-resolution photoacoustic
  overtone spectrum of acetylene near 570 nm using a ring-dye-laser
  spectrometer}, Chem. Phys. Lett. 180 (1991) 310 -- 316.
\newblock \href {http://dx.doi.org/10.1016/0009-2614(91)90325-4}
  {\path{doi:10.1016/0009-2614(91)90325-4}}.

\bibitem{13SiMeVa.C2H2}
M.~Siltanen, M.~Mets{\"a}l{\"a}, M.~Vainio, L.~Halonen, {Experimental
  observation and analysis of the $3\nu_1$ stretching vibrational state of
  acetylene using continuous-wave infrared stimulated emission}, J. Chem. Phys.
  139 (2013) 054201.
\newblock \href {http://dx.doi.org/10.1063/1.4816524}
  {\path{doi:10.1063/1.4816524}}.

\bibitem{83ScLeKl.C2H2}
G.~J. Scherer, K.~K. Lehmann, W.~Klemperer, {The high resolution visible
  overtone spectrum of acetylene}, J. Chem. Phys. 78 (1983) 2817--2832.
\newblock \href {http://dx.doi.org/10.1063/1.445269}
  {\path{doi:10.1063/1.445269}}.

\bibitem{13LyCaMo.C2H2}
O.~M. Lyulin, A.~Campargue, D.~Mondelain, S.~Kassi, {The absorption spectrum of
  acetylene by CRDS between 7244 and 7918 cm$^{-1}$}, J. Quant. Spectrosc.
  Radiat. Transf. 130 (2013) 327--334.
\newblock \href {http://dx.doi.org/10.1016/j.jqsrt.2013.04.028}
  {\path{doi:10.1016/j.jqsrt.2013.04.028}}.

\bibitem{14LyMoBe.C2H2}
O.~M. Lyulin, D.~Mondelain, S.~Beguier, S.~Kassi, J.~Vander~Auwera,
  A.~Campargue, {High-sensitivity CRDS absorption spectroscopy of acetylene
  between 5851 and 6341 cm$^{-1}$}, Mol. Phys. 112 (2014) 2433--2444.
\newblock \href {http://dx.doi.org/10.1080/00268976.2014.906677}
  {\path{doi:10.1080/00268976.2014.906677}}.

\bibitem{16KaLyBe.C2H2}
S.~Kassi, O.~M. Lyulin, S.~B\'eguier, A.~Campargue, {New assignments and a rare
  peculiarity in the high sensitivity CRDS spectrum of acetylene near 8000
  cm$^{-1}$}, J. Mol. Spectrosc. 326 (2016) 106--114.
\newblock \href {http://dx.doi.org/10.1016/j.jms.2016.02.013}
  {\path{doi:10.1016/j.jms.2016.02.013}}.

\bibitem{HITRAN2004}
L.~Rothman, D.~Jacquemart, A.~Barbe, D.~C. Benner, M.~Birk, L.~Brown,
  M.~Carleer, C.~Chackerian, K.~Chance, L.~Coudert, V.~Dana, V.~Devi, J.-M.
  Flaud, R.~Gamache, A.~Goldman, J.-M. Hartmann, K.~Jucks, A.~Maki, J.-Y.
  Mandin, S.~Massie, J.~Orphal, A.~Perrin, C.~Rinsland, M.~Smith, J.~Tennyson,
  R.~Tolchenov, R.~Toth, J.~V. Auwera, P.~Varanasi, G.~Wagner,
  \href{http://www.sciencedirect.com/science/article/pii/S0022407305001081}{The
  hitran 2004 molecular spectroscopic database}, Journal of Quantitative
  Spectroscopy and Radiative Transfer 96~(2) (2005) 139 -- 204.
\newblock \href {http://dx.doi.org/10.1016/j.jqsrt.2004.10.008}
  {\path{doi:10.1016/j.jqsrt.2004.10.008}}.
\newline\urlprefix\url{http://www.sciencedirect.com/science/article/pii/S00224%
07305001081}

\bibitem{HITRAN2008}
L.~S. Rothman, I.~E. Gordon, A.~Barbe, D.~C. Benner, P.~F. Bernath, M.~Birk,
  V.~Boudon, L.~R. Brown, A.~Campargue, J.~P. Champion, K.~Chance, L.~H.
  Coudert, V.~Dana, V.~M. Devi, S.~Fally, J.~M. Flaud, R.~R. Gamache,
  A.~Goldman, D.~Jacquemart, I.~Kleiner, N.~Lacome, W.~J. Lafferty, J.~Y.
  Mandin, S.~T. Massie, S.~N. Mikhailenko, C.~E. Miller, N.~Moazzen-Ahmadi,
  O.~V. Naumenko, A.~V. Nikitin, J.~Orphal, V.~I. Perevalov, A.~Perrin,
  A.~Predoi-Cross, C.~P. Rinsland, M.~Rotger, M.~Simeckova, M.~A.~H. Smith,
  K.~Sung, S.~A. Tashkun, J.~Tennyson, R.~A. Toth, A.~C. Vandaele,
  J.~Vander~Auwera, The hitran 2008 molecular spectroscopic database, J. Quant.
  Spectrosc. Radiat. Transf. 110 (2009) 553--572.

\bibitem{HITRAN2012}
L.~S. Rothman, I.~E. Gordon, Y.~Babikov, A.~Barbe, D.~C. Benner, P.~F. Bernath,
  M.~Birk, L.~Bizzocchi, V.~Boudon, L.~R. Brown, A.~Campargue, K.~Chance, E.~A.
  Cohen, L.~H. Coudert, V.~M. Devi, B.~J. Drouin, A.~Fayt, J.-M. Flaud, R.~R.
  Gamache, J.~J. Harrison, J.-M. Hartmann, C.~Hill, J.~T. Hodges,
  D.~Jacquemart, A.~Jolly, J.~Lamouroux, R.~J. {Le Roy}, G.~Li, D.~A. Long,
  O.~M. Lyulin, C.~J. Mackie, S.~T. Massie, S.~Mikhailenko, H.~S.~P.
  M{\"u}ller, O.~V. Naumenko, A.~V. Nikitin, J.~Orphal, V.~Perevalov,
  A.~Perrin, E.~R. Polovtseva, C.~Richard, M.~A.~H. Smith, E.~Starikova,
  K.~Sung, S.~Tashkun, J.~Tennyson, G.~C. Toon, V.~G. Tyuterev, G.~Wagner, {The
  {\it HITRAN} 2012 molecular spectroscopic database}, J. Quant. Spectrosc.
  Radiat. Transf. 130 (2013) 4 -- 50.
\newblock \href {http://dx.doi.org/10.1016/jqsrt.2013.07.002}
  {\path{doi:10.1016/jqsrt.2013.07.002}}.

\bibitem{jtlinTROVE}
K.~L. Chubb, S.~N. Yurchenko, A.~Yachmenev, J.~Tennyson, {TROVE: Treating
  linear molecule HCCH}, J. Chem. Phys.(to be submitted).

\bibitem{jt709}
C.~Sousa-Silva, L.~K. McKemmish, K.~L. Chubb, J.~Baker, E.~J. Barton, M.~N.
  Gorman, T.~Rivlin, J.~Tennyson, {Original Research By Young Twinkle Students
  (ORBYTS): When can students start performing original research}, Phys. Educ.

\end{thebibliography}

\end{document}